\documentclass[aps,pre,twocolumn,groupedaddress,floatfix,a4,showpacs]{revtex4-1}
\usepackage{amsmath,amssymb,bm}
\usepackage[version=3]{mhchem}
\usepackage[]{graphicx}
\usepackage{color}

\def\so3{\ce{SO3-} }
\def\hy {\ce{H3O+} }
\def\water {\ce{H2O}}

\def\h+{\ce{H+} }

\def\eg{{\it e.g.} }
\def\ie{{\it i.e.} }

\begin{document}
\title{
Morphology of supported polymer electrolyte ultra-thin films: a numerical study 
}
\author{Daiane Damasceno Borges}
\altaffiliation{Present address: Institut Charles Gerhardt Montpellier, UMR CNRS 5253, UM2, Place E. Bataillon, 34095 Montpellier Cedex 5, France} 
\email[]{daianefis@gmail.com}
\author{Gerard Gebel}
\affiliation{Univ. Grenoble Alpes, LITEN-DTNM, F-38000 Grenoble, France}
\affiliation{CEA, LITEN-DTNM, F-38000 Grenoble, France}
\author{Alejandro A. Franco}
\affiliation{Univ. Grenoble Alpes, LITEN-DTNM, F-38000 Grenoble, France}
\affiliation{CEA, LITEN-DTNM, F-38000 Grenoble, France}
\affiliation{Laboratoire de R\'eactivit\'e et Chimie des Solides (LRCS), CNRS UMR 7314, Universit\'e de Picardie Jules Verne, 80039 Amiens Cedex, France}
\affiliation{R\'eseau sur le Stockage \'Electrochimique de l'Energie (RS2E), FR CNRS 3459, France}
\author{Kourosh Malek}
\affiliation{Energy, Mining and Environment, National Research Council of Canada, Vancouver, BC, Canada}
\author{Stefano Mossa}
\email[]{stefano.mossa@cea.fr}
\affiliation{Univ. Grenoble Alpes, INAC-SPRAM, F-38000 Grenoble, France}
\affiliation{CNRS, INAC-SPRAM, F-38000 Grenoble, France}
\affiliation{CEA, INAC-SPRAM, F-38000 Grenoble, France}
\begin{abstract}
Morphology of polymer electrolytes membranes (PEM),~\eg Nafion, inside PEM fuel cell catalyst layers has significant impact on the electrochemical activity and transport phenomena that determine cell performance. In those regions, Nafion can be found as an ultra-thin film, coating the catalyst and the catalyst support surfaces. The impact of the hydrophilic/hydrophobic character of these surfaces on the structural formation of the films has not been sufficiently explored yet. Here, we report about Molecular Dynamics simulation investigation of the substrate effects on the ionomer ultra-thin film morphology at different hydration levels. We use a mean-field-like model we introduced in previous publications for the interaction of the hydrated Nafion ionomer with a substrate, characterized by a tunable degree of hydrophilicity. We show that the affinity of the substrate with water plays a crucial role in the molecular rearrangement of the ionomer film, resulting in completely different morphologies. Detailed structural description in different regions of the film shows evidences of strongly heterogeneous behaviour. A qualitative discussion of the implications of our observations on the PEMFC catalyst layer performance is finally proposed.
\end{abstract}
%
%\pacs{}
%
\date{\today}
\maketitle
\section{Introduction}
\label{sec:introduction}
The Membrane Electrode Assembly (MEA) is the core of a Proton Exchange Membrane Fuel Cell (PEMFC). It consists of two symmetric catalyst layers (CL), placed at the anode and cathode sides and separated by a polymer electrolyte membrane (PEM), and of the gas diffusion layer~\cite{Eikerling2007, Vielstich2003, Weber2004}. Despite the tremendous progresses achieved in the past decades, the PEMFC is not yet largely commercialized. The most significant hurdles for large scale production include reduction of costs, improvement of power density and enhancement of durability~\cite{Borup2007, Peighambardoust2010}. It is currently consensual that further development of PEMFCs implies a directly understanding of the material properties at the molecular level, for each component of the MEA. In particular, regions of crucial importance are the catalyst layers, where different electrochemical reaction mechanisms take place~\cite{Litster2004,Mehta2003}. This includes two half-cell reaction mechanisms: {\em i)} the Hydrogen Oxidation Reaction (HOR), \ce{H2 -> 2H+ + 2e-} at the anode; and {\em ii)} the Oxygen Reduction Reaction (ORR), \ce{O2 + 4H+ + 4e- -> 2H2O} at the cathode~\cite{Markovic2002, Damjanovic1967, DeMorais2011}. The rates of those reaction mechanisms determine the efficiency of electrochemical conversion, which is directly related to the fuel cell performance~\cite{Rinaldo2010,Franco2008}. The most efficient choice of catalyst particles for enhancing reaction rates are Pt-based particles. The high cost associated to the amount of platinum required for the catalyst, particularly at the cathode, is one of the drawbacks of fuel cells~\cite{Stamenkovic2007,Gasteiger2005,Eikerling2007b, Eikerling2009}.

The CL performance also depends on the transport conditions for reactants and products moving from (to) other MEA components from (to) the catalyst surface inside the CL. A good cathode CL performance (similarly for the anode CL) may depend on: transport of protons from the membrane to the catalyst; electron conduction from the current collector to the catalyst; reactant gases from gas channels to the catalyst; and correct removal of water from the catalyst layer~\cite{Eikerling2007b}. In order to meet all requirements, a complex structure with interconnected pores for reactants diffusion, a phase for electron conduction and a path for proton transport must be considered in devising a CL~\cite{Malek2007, Malek2011a, More2006,Xie2010}.

The necessity of having a heterogeneous structure to satisfy all catalyst layer functionalities, implies the quest for new materials design to optimize the distribution of transport media, in order to reduce transport losses and produce the highest current density with a minimum amount of catalyst particles~\cite{Litster2004}. Effective properties mainly depend on the nature of the materials used and fabrication process applied. During the preparation of catalyst layer ink, Pt/C agglomerates, Nafion ionomer and solvent are mixed together. This process is highly empirical and uses poorly controlled processing methods, which are not based on any knowledge of physico-chemical processes at the molecular level~\cite{Wilson1992, Wilson1993, Wilson1995}. 

Also, the CL is composed by materials characterized by very heterogeneous wetting properties, i.e., {\em hydrophilic} or {\em hydrophobic} character. The hydrophilicity of the CL plays an important role in fuel cell water management and it can be modified during the fabrication process~\cite{Li2009, Li2010}. Moreover, these wetting properties can be affected during fuel cell operation. The degradation mechanisms for these materials include ripening and compositional changes of catalyst due to corrosion, catalysts poisoning by adsorbed impurities, aging of the proton exchange electrolyte membrane, changes in the hydrophobic/hydrophilic properties of catalyst layer surfaces~\cite{Mashio2010, Borup2007,Chen2006, Wang2009}.

In Ref.~\cite{Borges2013} we introduced a mean-field-like model for the interaction of the hydrated Nafion ionomer with a substrate, characterized by a tunable degree of hydrophilicity. In particular, we focused on transport properties of water molecules in different regions of the film and demonstrated a high degree of heterogeneity. 
We also gave a few hints about the dependence of some morphological features on the wetting properties of the substrate~\cite{Borges2013, Borges2013b}. Here, we consider a much more extended set of simulation data and a provide a complete picture of the produced ultra-thin films morphology. We performed a comprehensive Molecular Dynamics (MD) computer simulation investigation of the substrate effects on the ionomer ultra-thin film morphology at different hydration levels, considering as the control parameter the hidrophilicity degree of the substrate. We have analyzed quantitatively morphology and topology of the films, both at the interfaces with the solid support and air, and in the central layers far from the boundaries. We propose a general qualitative scenario for thin-films morphology in different hydration conditions and wetting nature of the support. We finally speculate about possible implications of our work on the optimization of the actual devices.

The paper is organized as follows. In Section~\ref{sec:catalyst layer} we provide an overview of experimental and computer simulation work relevant in the present context. In Sect.~\ref{sec:ionomer-model} we describe the atomistic model used for mimicking the hydrated ionomer and our effective model for the interaction of the ionomer with the substrate. We characterize the wetting properties of the support in terms of a contact angle. We finally give a few details on our computer simulations scheme. More technical details can be found in the Supplementary Information accompanying this paper. In Sect.~\ref{sec:morphology} we report our extended investigation of the morphology, while in Sect.\ref{sec:PEMFC-technology} we focus more in details on both the support/ionomer and ionomer/vacuum interfaces, discussing the implications of our findings on PEMFC technology. Finally, Sect.~\ref{sec:conclusions} contains our conclusions and possible perspectives on further work.

\section{The catalyst layer}
\label{sec:catalyst layer}
The CL structure is formed by platinum nanoparticles dispersed on a carbon matrix with impregnated Nafion ionomer~\cite{Malek2007, Malek2011a, More2006,Xie2010}. Nafion is a perfluorinated polymer which results from the copolymerisation of a tetrafluoroethylene backbone (Teflon) and perfluorovinyl ether groups, terminated by sulfonate group side-chains~\cite{Moore2004}. Nafion is characterized by a highly heterogeneous structure at the nanoscale, due to a spontaneous phase separation of the hydrophobic backbones and hydrophilic sulfonated side chains upon hydration~\cite{Gierke1981, Hsu1983,Yeager1981,Gebel1987, Gebel2000a,Young2002,Rubatat2002,Schmidt-Rohr2008,Elliott2011}. Nafion has been introduced as one of CL constituents for two reasons~\cite{Litster2004}: first, during the fabrication process it acts as a binder, playing an important role on the dispersion of Pt/C aggregates and, as a consequence, on the Pt utilization. Second, during fuel cell operation, it forms an extended proton-conductor network available for proton migration from (to) the membrane to (from) the catalyst sites. Nafion inside CL presents an inhomogeneous and non-continuous phase. It can be found as a well-dispersed ultra-thin film on the surface of carbon supports and Pt particles. Typically, this film is not uniformly distributed and has a thickness spanning the range $\sim 4$ to $20$ nm~\cite{More2006}.

The formation of Nafion ultra-thin films inside the catalyst layer has been analysed in numerous recent  studies~\cite{Ma2007, Paul2011, Paul2011a, Paul2013, Wood2009, Dura2009, Masuda2009, Koestner2011, Eastman2012, Nagao2013, Kusoglu2014, Modestino2012, Modestino2013}. Structure and properties of these films significantly differ from those in the ionomer membrane (bulk). A detailed study based on variation of the ionomer film thickness and comparison with the membrane, has shown that some ionomer properties, {\em e. g.}, water uptake, swelling, water diffusion, respond differently to relative humidity. There is a critical thickness of around $60$~nm, where a transition from a bulk-like to confined ionomer is observed~\cite{Eastman2012}. Other experiments in thin-films adsorbed on \ce{Si2O}-terminated surfaces have underlined a proton conductivity which is lower than in the case of the bulk membrane~\cite{Paul2011,Paul2011a}. Also, Atomic Force Microscopy (AFM) experiments have shown that the ionomer orientation depends on the atomic arrangement of the substrate surface~\cite{Masuda2009,Koestner2011}. In the CL the Nafion ionomer is expected to self-organize in different forms, depending on the properties of the substrate. The impact of surface hydrophilicity on the ionomer properties have been recently subject of many studies, and there is experimental evidence that the change of wetting properties of the substrate is sufficient to affect Nafion film morphology~\cite{Modestino2012, Modestino2013, Bass2010, Bass2011}. 

Modestino {\em et al.}~\cite{Modestino2012} have investigated the possibility to control structure and properties of Nafion thin films by modifying the wetting properties of the substrate. They prepared Nafion thin-films deposited on hydrophobic (OTS passivated Si) and hydrophilic (silicon) substrates, and investigated the impact of the internal morphology on water uptake. They found that thin films cast on hydrophobic substrates result in parallel orientation of ionomer channels, which retards the absorption of water from humidified environments. In contrast, films prepared on \ce{SiO2} result in isotropic orientation of these domains, thus favoring water adsorption and swelling of the polymer. 

Wood {\em et al.}~\cite{Wood2009} observed multilayer structures of Nafion thin films in contact with smooth flat surfaces. These structures consist of separate hydrophobic and hydrophilic domains formed within the Nafion layer, when equilibrated with saturated \ce{D2O} vapor. Any strong interaction between a flat surface and Nafion is likely to lead to the polymer chains lying flat on that surface, which is completely different from any bulk Nafion morphologies proposed so far. When Nafion was in contact with a bare Pt surface, a hydrophobic Nafion region was found to form adjacent to a Pt film. In contrast, when a PtO monolayer was present, the hydrophobic backbone was pushed outward and the hydrophilic side chains were in contact with the PtO surface. These restructuring processes were reversible and strongly influenced by the polymer hydration. Dura {\em et al.}~\cite{Dura2009} performed Neutron Refractometry (NR) measurements in order to investigate the structure of Nafion in contact with \ce{SiO2}, \ce{Au} and \ce{Pt} surfaces. They showed that lamellar structures, composed of thin alternating water-rich and Nafion-rich layers, exist at the interface between \ce{SiO2} and the hydrated Nafion film. However, multilamellar structures do not exist at the Pt/Nafion or Au/Nafion (metallic) interfaces, where a single thin layer rich in water occurs. This difference indicates that Au and Pt surfaces have a lower affinity to the sulfonic acid/water phase than the more hydrophilic \ce{Si2O} surface. These structures where interpreted in terms of an interface-induced ordering of the ribbon-like aggregates or lamellae observed in Small-Angle X-Rays Scattering (SAXS) experiments of bulk Nafion. Therefore, the first Nafion-rich layer could be formed by closely packed ribbons or lamellae, oriented with their faces parallel to the substrate, and with successive layers of increasingly disordered character. 

Molecular dynamics (MD) simulations can also provide insights in clarifying nanoscale structure and transport properties of Nafion at interfaces. Despite this evidence, only a few numerical studies have been dedicated to the above issues, partly due to the issue of convincingly parametrizing interaction force fields between Nafion and substrate materials. A few examples are reported in what follows.

Most part of computational work has focused on the behaviour of Nafion in the presence of carbon and platinum based materials~\cite{Balbuena2005,Lamas2006,Liu2008,Selvan2012,Selvan2008}. These simulations showed that Nafion strongly interacts with Pt nanoparticles, mainly through the hydrophilic sulfonic chains. Mashio {\em et al.}~\cite{Mashio2010} analysed the morphology of Nafion ionomer and water in contact with graphite surfaces. Because of the hydrophobic nature of the graphite sheet and ionomer backbones, Nafion ionomer was found to interact with the graphite sheet mainly via the backbones, whereas side chains were  oriented away from the graphite sheet and water molecules were adsorbed at the sulfonic acid groups. The effect on structure and transport properties of the functionalization of graphitized carbon sheet with carboxyl ($COOH$) or carboxylate ($COO^-$) groups was also explored. The most significant effect on water and ionomer distributions was shown to come from the graphite sheet functionalized with ion groups. It was observed that the number of water molecules, hydronium ions and sulfonic acid groups in the vicinity of the graphite sheet increases with the application of the ionized functional groups. Overall, the structure and surface properties of carbon supports clearly affect the transport properties of proton and water. 

\section{Modelling}
\label{sec:ionomer-model}
\subsection{The ionomer model}
\label{subsec:ionomer-model}
The Nafion polymer, is formed by a hydrophobic polytetrafluorothylene backbone ([\ce{-CF_2-CF_2}]) and intercalated perfluorinated side-chains, which are terminated by a strongly hydrophilic radical sulfonic acid group ($\ce{SO_3H}$). We consider a united-atom representation for $\ce{CF}$, $\ce{CF_2}$ and $\ce{CF_3}$ and a fully atomistic model for the \so3 groups in the side-chain~\cite{Urata2005}. This mixed modelling scheme is commonly used to represent Nafion~\cite{Allahyarov2007,Allahyarov2009,Cui2007,Cui2008,Vishnyakov2000,Vishnyakov2001,Liu2010}. 
The polymer backbone is formed by a linear chain of $160$ bonded monomers, which corresponds to a (completely extended) length of approximately $24$~nm. $10$ side-chains are uniformly distributed along the backbone. Each side-chain has $11$ atoms and a length of approximately $1$~nm. The spacing between adjacent side-chains has been chosen in order to match an equivalent weight $\sim 1143.05$~g/mol of $\ce{SO_3^-}$, a value typical for commercial Nafion 117. 

The simulation starts from a configuration created by randomly placing 20 polymer chains, 200 hydronium ions and the number of water molecules set according to the desired water content $\lambda$. The system was equilibrated after a series of annealing and optimization runs. After the equilibration, trajectories of, at least, $5$~ns were generated for analyses. The total interaction energy of the system is the sum of non-bonded and intramolecular bonded terms. The force field parameters of our model are similar to the ones of the fully atomistic model of Venkatnathan~\cite{Venkatnathan2007} and adapted to the united-atom representation. The polymer backbone is charged neutral, while the sulfonic acid head groups are assumed to be fully ionized (\so3). In order to preserve charge neutrality, flexible hydronium complexes (\ce{H3O+}) were added, with force field parameters and partial charges taken from~\cite{Kusaka1998}. Water molecules are described by the rigid extended Simple Point Charge (SPC/E) model~\cite{Berendsen1987}. A list of all parameters is given in Table~1 in the Supporting Information. 
We tested the reliability of the ionomer model by performing various simulations of hydrated Nafion in the bulk and compared our results with those found in the literature. Our model is able to reproduce the general Nafion morphology and the correct dynamics of water and hydronium ions. For the reader interested, the main results of Nafion membrane are reported in Supporting Information.
\begin{figure}[t]
\centering
\includegraphics[width=0.45\textwidth]{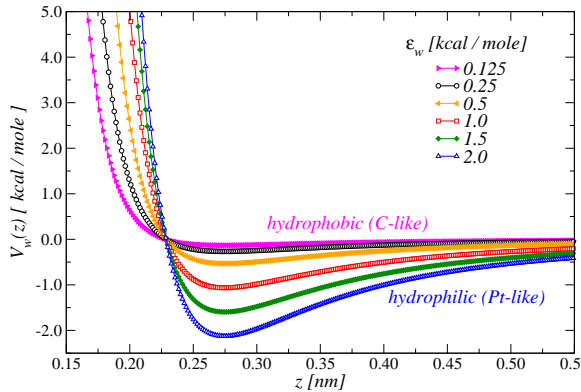}
\includegraphics[width=0.45\textwidth]{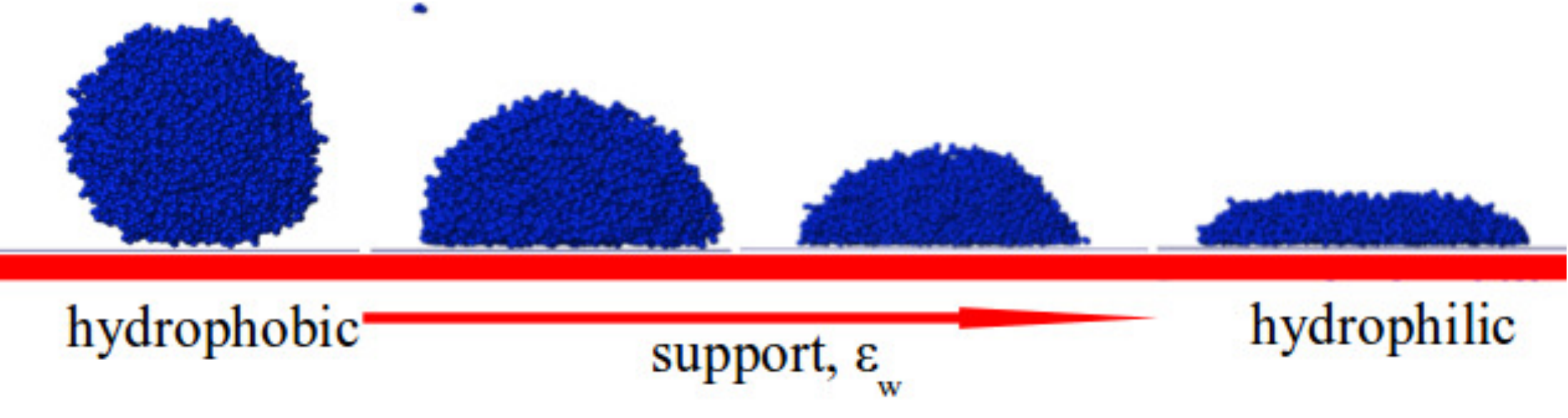} 
\caption{
{\em Top:} 9-3 Lennard-Jones potential function for different values of $\epsilon_{w}$, with $\sigma_{w}=0.32$~nm. Most hydrophilic case on the bottom. {\em Bottom:} Simulated clusters formed by $3500$ water molecules in contact with the support characterized by increasing values of $\epsilon_{w}$. It is evident the increasingly hydrophilic nature of the interaction.
}
\label{fig1:93lj}
\end{figure}

\subsection{The interaction with the support} 
\label{subsec:support-model}
The effect of confinement due to the presence of a solid phase characterized by given wetting properties is mimicked by the interaction potential of the ionomer with the support. The hydrophobic or hydrophilic character of a surface is related to nano-scale features, such as structure and polarity~\cite{Giovambattista2007,Castrillon2009,Nijmeijer1990}. Here we have considered a mean-field-like interaction ionomer/substrate, that allows us to precisely control the hydrophilic character of the substrate by using a unique tunable control parameter. This strategy has already been successfully applied in studies of molecular liquids at interfaces, like pure water in contact with perfectly smooth walls~\cite{Scheidler2002,Spohr1988}. All system units interact with an infinite smooth unstructured wall (the support), placed at $z=0$ and parallel to the $xy$-plane, {\em via} a $9\text{--}3$ Lennard Jones potential~\cite{Abraham1977}. This only depends on the distance, $z$, of the unit from the support:
\begin{equation}
V_{w}^\alpha(z)=\epsilon^\alpha_w\left[ \frac{2}{15}\left(\frac{\sigma_w^\alpha}{z}\right)^9-\left(\frac{\sigma_w^\alpha}{z}\right)^3\right]\theta(z_c-z),
\label{eq:wall}
\end{equation}
where $z_c=$~1.5 nm is a cut-off distance and $\theta$ is the Heaviside function. 
\begin{table}[t]
\footnotesize
\centering
\begin{tabular}{l r r r r r r}
\hline\hline 
$\epsilon_{w}$ (kcal/mole) &  0.125  & $^*$0.25 & 0.5 & $^*$1.0    & $^*$1.5	 & $^*$2.0	\\ 
\textbf{$\theta$} (degrees) & $163.0$ &  $151.3$ & $136.3$ & $100.9$ &	$69.1$  & $29.7$  \\
\hline\hline
\end{tabular}
\caption{Values of the water droplet contact angles at the indicated values of $\epsilon_{w}$. We indicate with $^*$ the values of $\epsilon_{w}$ which we will consider in our analysis of the supported thin-films.}
\label{tb:contactangle}
\end{table}
The index $\alpha$ identifies complexes ($H_2O$, $H_30^+$, $SO_3^-$) with significant dipolar coupling to the (hydrophilic) support ($\alpha=\text{phyl}$), or units corresponding to the hydrophobic sections of the polymer ($\alpha=\text{phob}$) which, in contrast, interact very mildly. The energy well $\epsilon_w^\text{phob}=0.5$~kcal/mole is fixed and is the typical strength of the interaction of polymer units with a carbon sheet. This choice is justified by the observation that chemical and physical processes occurring at the surface, \eg adsorption and chemical reactions in operating PEMFC, can affect surface polarity~\cite{Mashio2010, Giovambattista2007}. These polarity changes do not affect the interaction with the (apolar) backbone monomers in the same way they modify the interaction with water molecules. The impact of the polarity of the substrate is therefore expected to be more important on the wall/water than on wall/ionomer interactions. The hydrophilicity parameter $\epsilon_w^\text{phyl}=\epsilon_w$ is the control parameter, which was systematically varied in the range $0.125$ to $2.0$~kcal/mol. The typical interaction length scale $\sigma_w^\alpha=0.32$~nm in all cases. Examples of the potential of Eq.~(\ref{eq:wall}) at the indicated values of $\epsilon_w$ are shown in Fig.~\ref{fig1:93lj} (top).
\begin{figure}[t]
\centering
\includegraphics[width=0.45\textwidth]{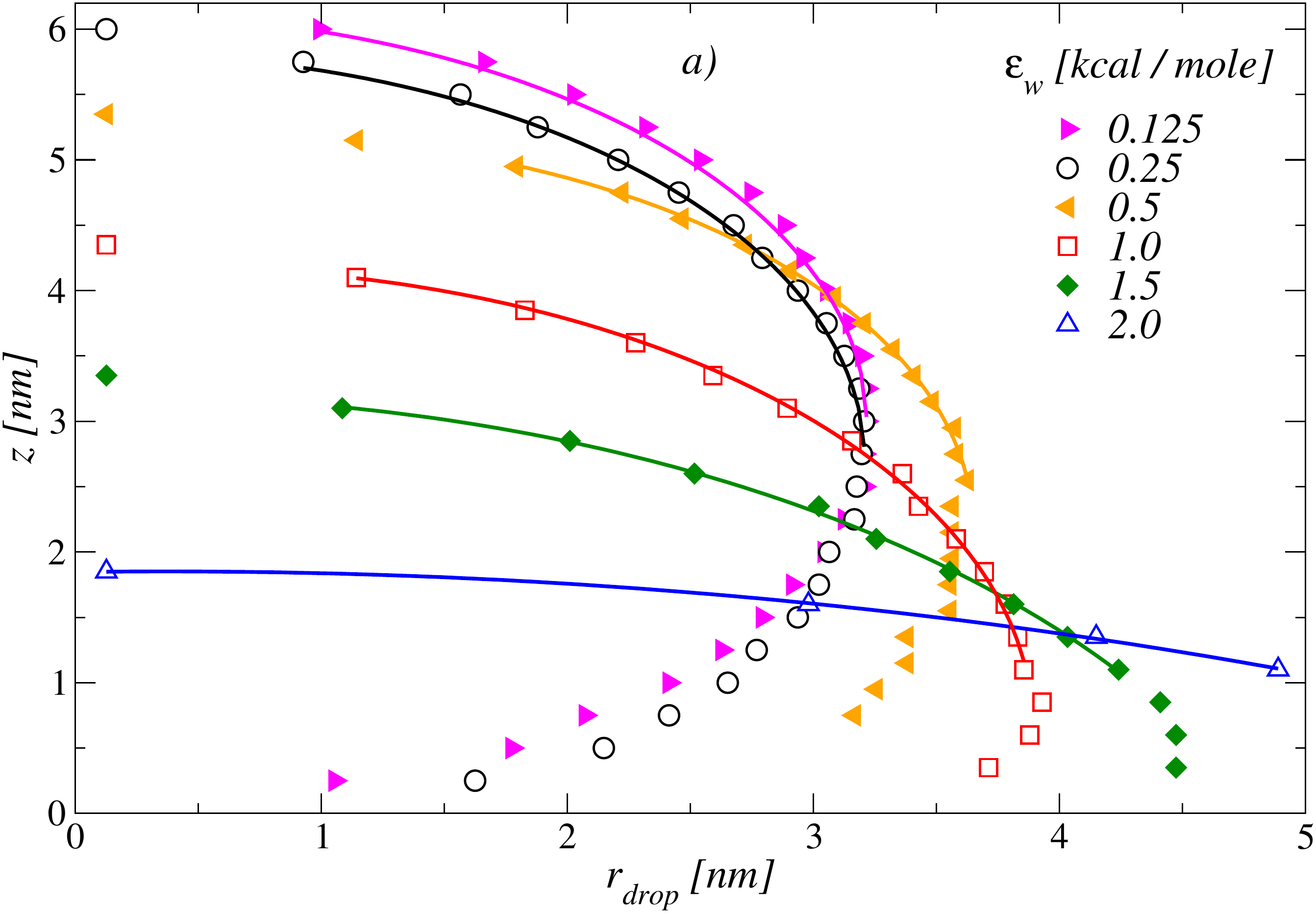}
\includegraphics[width=0.45\textwidth]{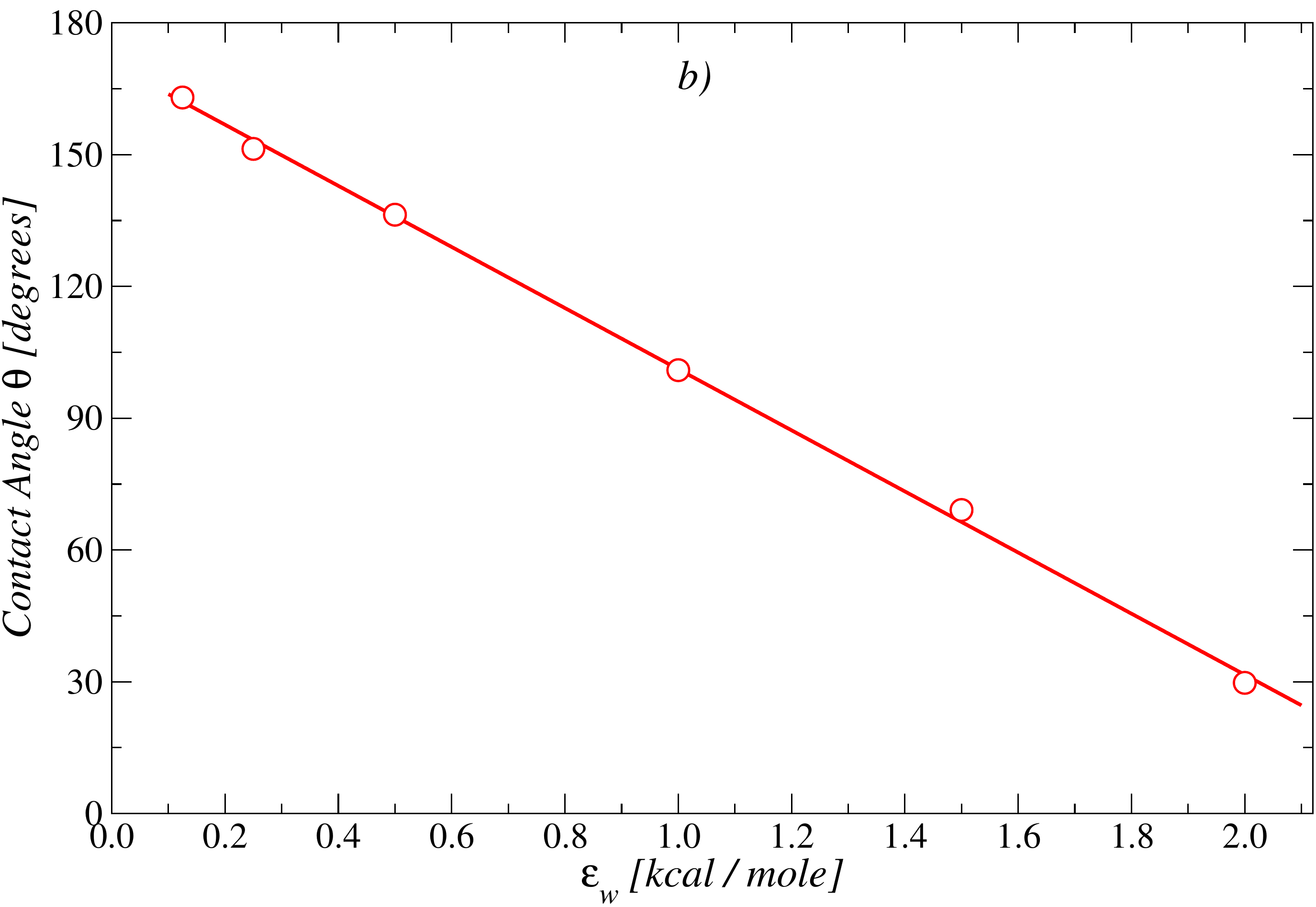}
\caption{
{\em a)} Water droplet profiles at the indicated values of $\epsilon_{w}$. The solid lines are the results of the fitting procedure discussed in the text. {\em b)} Contact angles extracted from the droplet profiles. $\theta$ varies linearly in the investigated $\epsilon_{w}$ range. 
}
\label{fig2:dropprofile}
\end{figure}
 
\begin{figure*}[]
\centering
\includegraphics[width=0.96\textwidth]{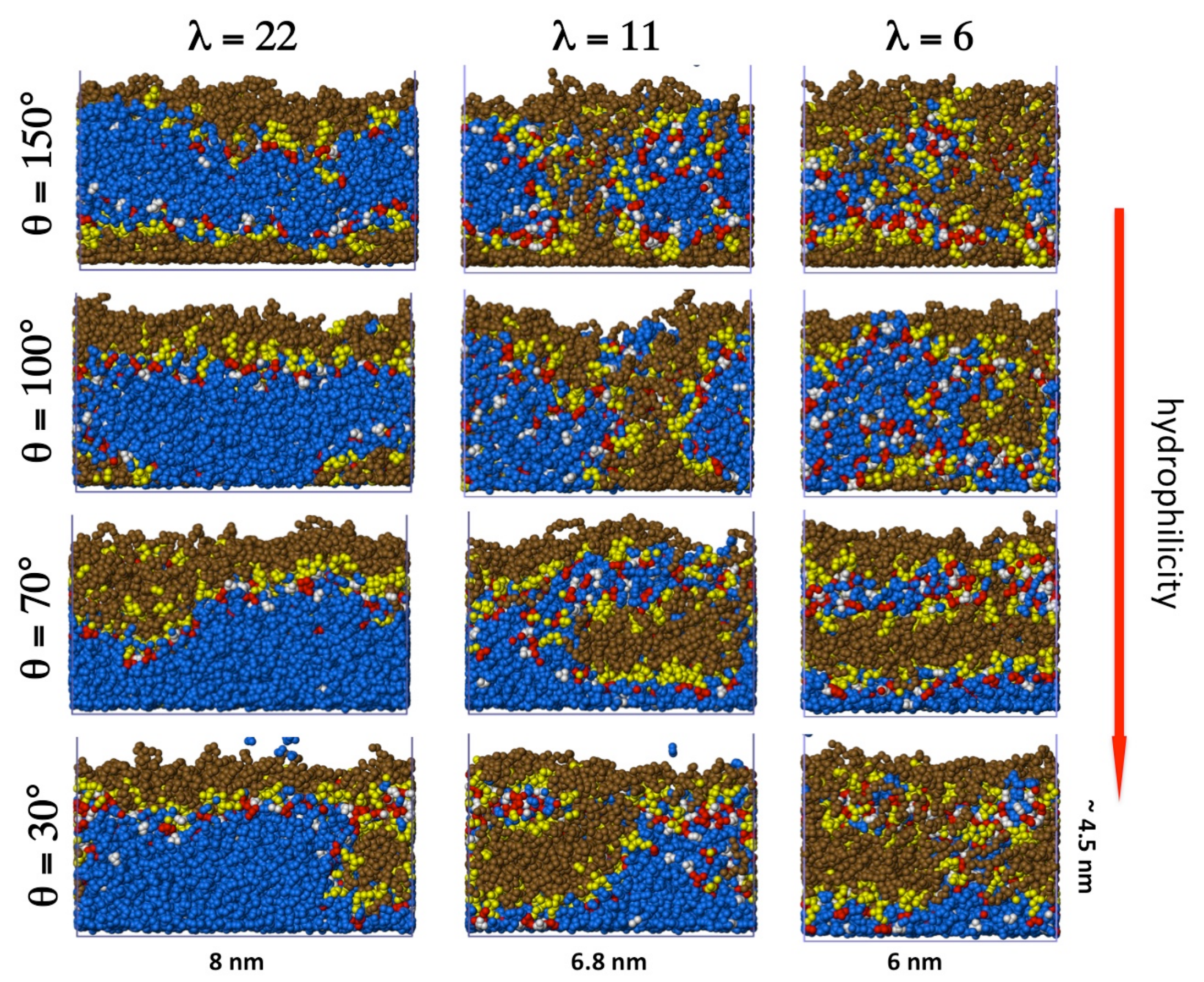}
\caption{
Lateral views of typical snapshots of hydrated Nafion thin-films at $\lambda=22$, 11 and 6, formed in contact with supports at the indicated values of the contact angle. These range from strongly hydrophobic ($\theta=150^\circ$) to very hydrophilic nature ($\theta=30^\circ$). The typical films thickness is about 4.5~nm. Beads pertaining to backbones are shown in brown, those pertaining to side-chains are in yellow, \ce{SO_3} groups are in red, water molecules in blue and hydronium ions in white.
}
\label{fig3:side}
\end{figure*}

\subsection{Wetting properties of the support and water droplet contact angles} 
\label{subsec:contact-angle}
In order to associate a physical meaning to the adopted choice for $\epsilon_{w}$ we have performed additional simulations of water droplets gently deposited on supports described by Eq.~(\ref{eq:wall}) and calculated the corresponding contact angles, $\theta$. By convention, a value of $\theta\le\pi/2$ corresponds to an hydrophilic support, while $\theta>\pi/2$ to an hydrophobic one. 

Figure~\ref{fig1:93lj} (bottom) shows typical snapshots of the equilibrated water droplets at the values $\epsilon_{w}=$ 0.25, 1.0, 1.5 and 2.0~kcal/mol. Already from visual inspection, the increasing hydrophilic character of the support is evident. The contact angles can next be estimated by fitting the droplet profiles~\cite{Shi2009,Werder2003}. Droplets profiles for different values of $\epsilon_{w}$ are shown in Fig.~\ref{fig2:dropprofile}~(a). A circular best fit through these points is extrapolated to the wall surface and provides $\theta$. We compute $\theta$ for each value of $\epsilon_{w}$. In Fig.~\ref{fig2:dropprofile}~(b) we plot $\epsilon_w$-dependence of the contact angle, which is linear in the investigated range. The associated contact angles to the $\epsilon_{w}$ are displayed in Table~\ref{tb:contactangle}. We will often refer to these values in what follows.

Altogether, these data prove that our strategy is able to provide us with different scenarios for the wetting character of the substrate, ranging from strongly hydrophobic to very hydrophilic conditions. Note that these values are representative of specific materials studied in the past. For example, computer simulations of water droplets on a platinum surface shows a contact angle $\theta\simeq$~20-30$^\circ$ \cite{Shi2009}. In the case of carbon nanotubes, the contact angle varies in the range 100$^\circ$ to 106$^\circ$, while for graphite from 110$^\circ$ to 115$^\circ$~\cite{Werder2003, Werder2001}. 

\section{Morphology of the hydrated ionomer thin-films}
\label{sec:morphology}
In Fig.~\ref{fig3:side} we show typical snapshots of the self-organized ionomer thin-films at the indicated values of hydration level and contact angles. Four hydrophilicity levels have been considered, encompassing very hydrophobic $(\theta \approx 150^\circ)$, intermediate $(\theta \approx 100^\circ)$, hydrophilic ($\theta \approx 70^\circ$) and strongly hydrophilic $(\theta \approx 30^\circ)$ supports. These contact angles correspond to interaction energies $\epsilon_{w}=$ 0.25, 1.0, 1.5 and 2.0~kcal/mol respectively, as detailed in Table~\ref{tb:contactangle}. The water contents considered are 6, 11 and 22. Those values are typical hydration level found in electrodes in fuel cell operation. Side-chains (yellow beads) terminated by the \so3 groups (red beads), decorate the interface between the backbone (brown beads) and the hydrophilic domains formed by water molecules and hydronium ions (blue and white beads, respectively). This configuration is typical of the phase-separated structure present in the Nafion membrane (bulk). The films thickness is about 4.5~nm, for all cases. By visual inspection, it is clear that the hydrophilicity of the substrate indeed controls the global morphology of the film. Also, it is evident that morphology and connectivity of the hydrated domains within the film, changes significantly  at different values of $\theta$ and $\lambda$. In what follows we report our analysis work and quantify these changes.

\subsection{Mass density distributions} 
\label{subsec:mass density}
\begin{figure}[t]
\centering  
\includegraphics[width=0.5\textwidth]{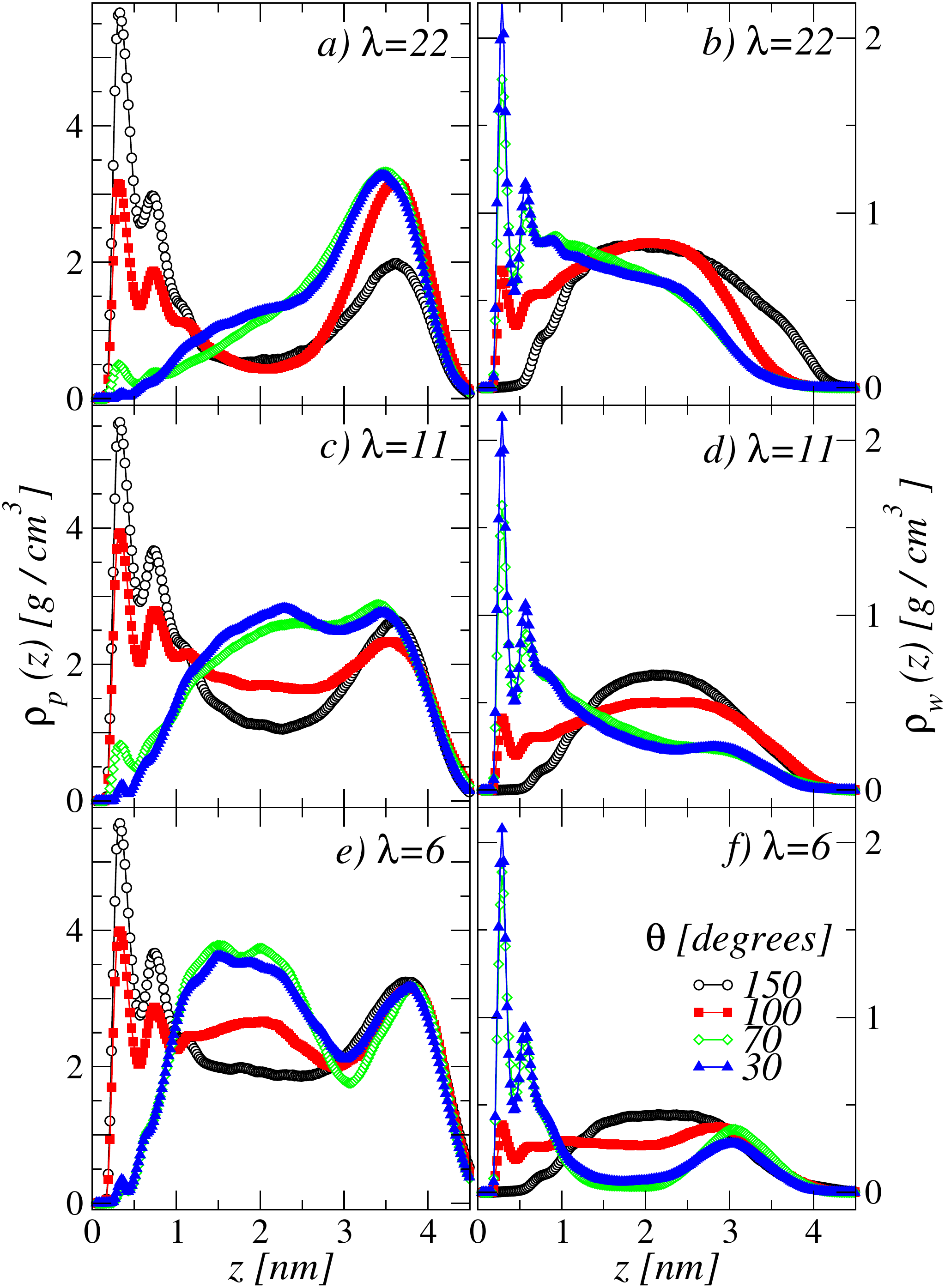}
\caption{
Mass density profiles for polymer ($\rho_{p}(z)$) and water molecules ($\rho_{w}(z)$) in the considered thin-films at $\lambda = 22$, 11 and 6 at the indicated values of the contact angles $\theta$. $z$ is the distance from the support.
}
\label{fig4:densityProfile}
\end{figure}

The structure of the ionomer film is first analysed in terms of the mass density profiles along the $z$-direction, perpendicular to the substrate. In Fig.~\ref{fig4:densityProfile} we show the polymer (left) and water (right) mass density distributions, $\rho_{p}(z)$ and $\rho_w(z)$ respectively, corresponding to snapshots of Fig.~\ref{fig3:side}. These curves clearly show important complementary changes on the distributions of water and polymer, following the value of $\theta$. 

We first focus on films on top of strongly hydrophobic surfaces ($\theta=150^\circ$). In the highly hydrated film ($\lambda=22$), at short distances from the surface, {\em i.e.} $z<1$~nm, the presence of polymer is dominant, while $\rho_{w}(z)$ shows almost no presence of water molecules at distances $z<0.5$~nm (Fig.~\ref{fig4:densityProfile} (a) and (b)). In this region, $\rho_{p}(z)$ presents two well defined peaks. In the center of the film, {\em i.e.} at distances $1.0<z<3.0$~nm, $\rho_{w}(z)$ is at its maximum value, while $\rho_{p}(z)$ is at the minimum. This suggests the formation of water domains confined between polymer-rich layers localized at the bottom and on top of the film. When decreasing the degree of hydration ($\lambda=$~11 and 6) this layered structure is less evident and the distribution of the polymer is less localized. As indicated in Fig.~\ref{fig4:densityProfile}~(e) the polymer density profile only has a shallow minimum in the latter case. 

In the case $\theta=100^\circ$, one starts to observe the presence of water molecules in direct contact with the substrate, as shown by the appearance of a peak in $\rho_{w}(z)$ at very short $z$. This suggests that a threshold exists at a value of the contact angle included in the range $100^\circ\div 150^\circ$, marking a transition from a completely hydrophobic to a mixed hydrophilic/hydrophobic character. In contrast, the polymer density profile shows the intensity of the first peaks are substantially decreased. Therefore, once water molecules start to adsorb at the support, the ionomer self-organizes by increasingly moving upward, and both species populate the substrate. With decreasing $\lambda$ this equilibrium is altered and the presence of polymer on the substrate is still dominant.

In the more hydrophilic cases ($\theta=70^\circ$ and $30^\circ{}$), the fraction of polymer in direct contact with the substrate is strongly reduced. At $\lambda=22$, the presence of ionomer is significant only for distances $z>2.5$~nm due to the presence of a large amount of water on the bottom which pushes the polymer upward, forming an ionomer layer on the upper part of the film. When $\lambda$ is lowered to 6, a significant fraction of the ionomer can be already found at a distance $z\simeq$~1~nm (Fig.~\ref{fig4:densityProfile}~(e)). In contrast, almost no water molecules are found in the middle of the film, in the range $1.0<z<2.5$~nm. This range encompasses the broad peak characterizing the polymer distribution and water molecules are concentrated in the region corresponding to a minimum of the polymer density profile.
\begin{figure}[t]
\centering  
\includegraphics[width=0.49\textwidth]{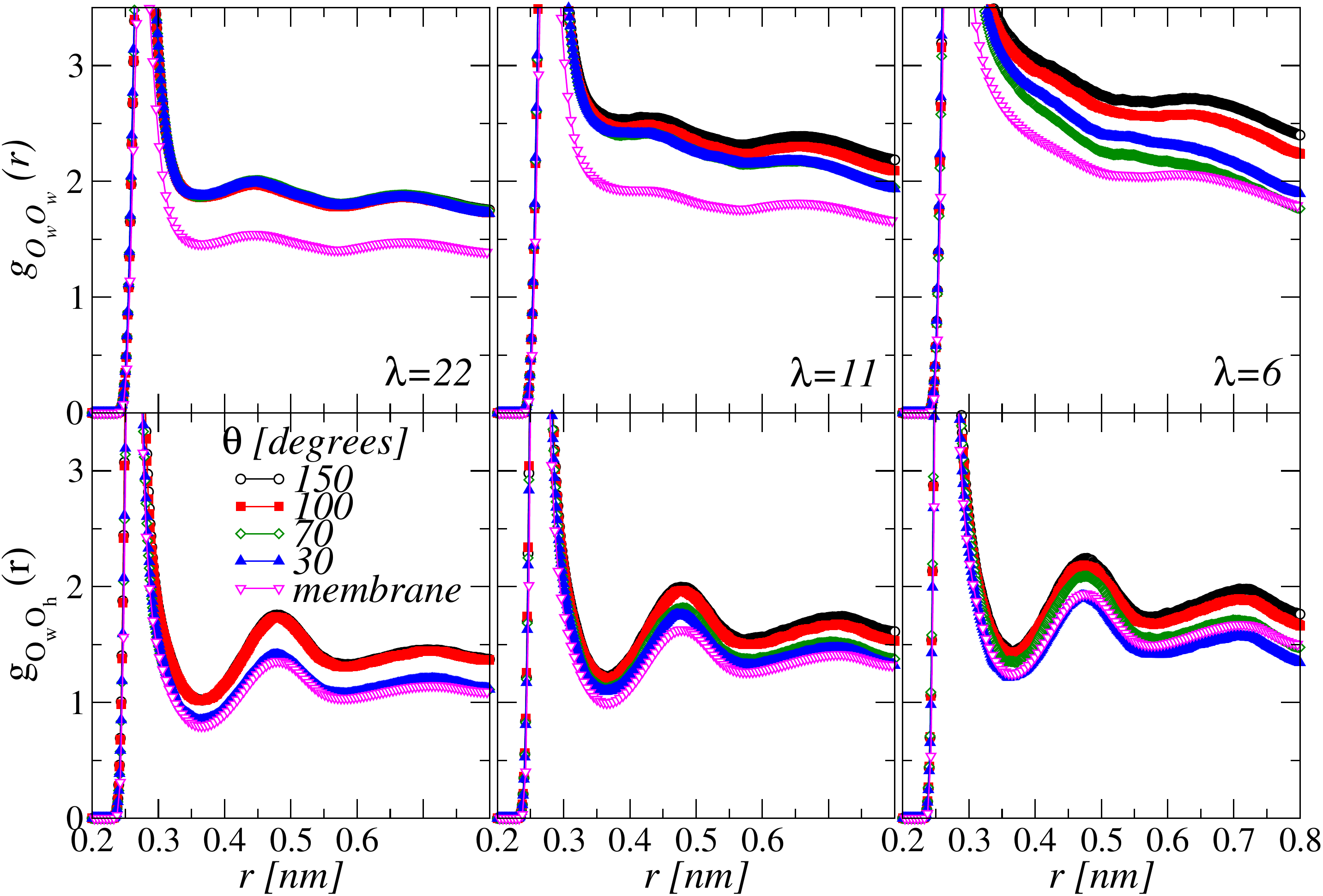}
\caption{
Radial distribution functions calculated from water/water oxygen atoms ($g_{O_wO_w}(r)$) and water/hydronium oxygen atoms ($g_{O_wO_h}(r)$) at $\lambda=$~22, 11 and 6 and at the indicated values of $\theta$. Data for the membrane in the same hydration conditions are also shown, for comparison.
}
\label{fig5:gOw-film}
\end{figure}

For all cases the positions of the two peaks in the vicinity of the wall for both $\rho_p$ and $\rho_w$ (at 0.29 and 0.55~nm for water, and 0.33 and 0.76~nm for polymer, respectively) do not change neither with hydration nor with surface hydrophilicity. The positions of those peaks are directly controlled by the interaction of the chemical units with the wall and, more precisely, by the parameter $\sigma_{w}=$~0.32~nm in Eq.~(\ref{eq:wall}). The relative distances between the two peaks (0.26~nm and 0.46~nm) are comparable with the nearest-neighbours distances between water molecules and between polymer beads and other species, respectively. Also, the oscillations in density profile (layering) are a typical feature of liquids at the interface with smooth walls~\cite{Spohr1989, Lee1994}.

From the above analysis we can conclude that the modulation of the interaction with the support has indeed a strong impact on local density profiles and, as a consequence, on the morphology of the thin-films. Although it is not surprising that the support wetting behaviour grows due to an increasing hydrophilic character, the overall density profiles are complex and extremely variable. A deeper understanding of the morphological features of these thin-films implies a more detailed analysis, that we will discuss in what follows.

\subsection{Radial Distribution Functions}
\label{subsec:gdr}
In this Section, we explore in details the local structure of the thin-films in terms of 3-dimensional partial radial distribution functions, $g_{\alpha\beta}(r)$, between selected chemical species $\alpha$ and $\beta$, for all the investigated systems. The $g_{\alpha\beta}(r)$ are properly normalized to the entire film volume. 

Fig.~\ref{fig5:gOw-film} shows the $g_{\alpha\beta}(r)$ for the oxygen atoms pertaining to water/water ($g_{O_wO_w}(r)$) and water/hydronium ($g_{O_wO_h}(r)$). We observe that the positions of the peaks are very similar to those for the membrane, while the intensity of the peaks, decreases when increasing the hydrophilicy of the substrate. The fist coordination number of water molecules around hydronium ions is reduced. For the case of $\lambda=22$, it decreases from 4.37 for $\theta=150^\circ$ and in the bulk, to 3.66 for $\theta=30^\circ$, indicating that a smaller number of water molecules is found in the vicinities of hydronium ions for the films formed on most hydrophilic supports.
\begin{figure}[t]
\centering  
\includegraphics[width=0.45\textwidth]{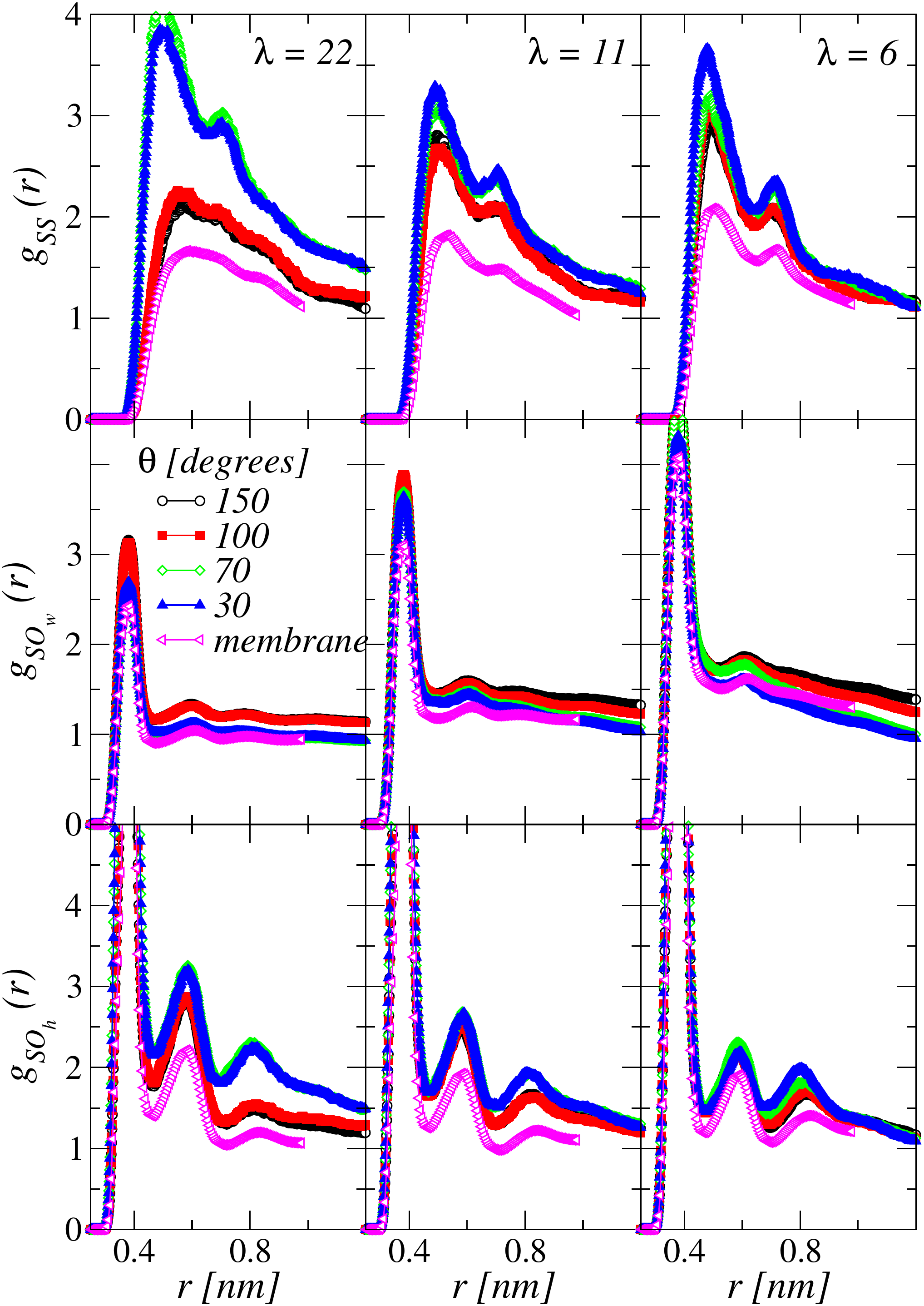}
\caption{
Radial distribution functions for sulphur-sulphur ($g_{SS}(r)$), suphur/water  ($g_{SO_w}(r)$), and sulphur/hydronium  ($g_{SO_h}(r)$) for $\lambda=$~22, 11 and 6, at the indicated values of $\theta$. Data for the membrane are also shown, for comparison.}
\label{fig6:gSS-film}
\end{figure}

The local structure around the \ce{SO3-} groups is investigated considering the $g_{\alpha\beta}(r)$ of sulphur atoms with sulphur, $g_{SS}(r)$, and water, $g_{SO_w}(r)$, and hydronium, $g_{SO_h}(r)$, oxygen atoms. These data are shown in Fig.~\ref{fig6:gSS-film}. At variance with the cases of water and hydronium discussed above, the $g_{SS}(r)$ calculated for the different films are very different when compared to the bulk case. This effect is accentuated at $\lambda=22$ (Fig.~\ref{fig6:gSS-film}). For $\theta=30^\circ$, the first peak is located at $0.49$~nm and an additional peak exists at $\simeq 0.7$~nm. When the hydrophilicity degree decreases, for $\theta=100^\circ$ and $150^\circ$, the first peak is shifted to $0.58$~nm, while the second one transforms into a shoulder, approaching the structureless $g_{SS}(r)$ found in the membrane. This indicates that the ionomer formed on a hydrophilic support self-organizes in such a way to have the \ce{SO3-} groups at distances smaller than those found for more hydrophobic cases or in the membrane. Consequently the number of \ce{SO3-} ions lying together is larger in the case of $\theta=30^\circ$. A possible conclusion is that for highly hydrated films ($\lambda=22$) the interaction of the film with the substrate transforms a bulk-like local structure, where \ce{SO3-} groups are less constrained and more spaced, into a configuration where the \ce{SO3-} groups form compact ionic domains. 

Both $g_{SO_w}(r)$ and $g_{SO_h}(r)$ exhibit strong correlations, similar to what is observed in the bulk (Figs.~\ref{fig6:gSS-film}). The first and second peaks are observed around $0.38$ and $0.60$~nm and these positions do not vary with the hydrophilicity of the support or with the hydration level of the film. Only the amplitude of those peaks show some changes with $\theta$ and $\lambda$. From the first shell coordination number of water molecules and hydronium ions around the sulphur atoms, we found that the number of water molecules surrounding the \ce{SO3-} decreases when the hydrophilicity of the substrate increases, while the opposite trend is observed for the hydronium. As it could be expected, these changes are more evident at $\lambda=22$, with water and hydronium coordination numbers varying respectively from $6.01$ and $1.45$ in the hydrophilic case, to $6.94$ and $0.9$ in the hydrophobic case. These findings are consistent with the picture based on the $g_{SS}(r)$ data. The number of water and hydronium molecules around the sulphur atoms is always correlated with the \ce{SO3-} agglomeration. Indeed, when the sulfonate ions are less agglomerated, they leave more space available for the water molecules to come closer to \ce{SO3-} groups. Consequently, the hydronium ions are increasingly solvated.

In summary, we have observed that for $\theta=30^\circ$ and $70^\circ$ sulphur atoms are found in compact agglomerates. As a consequence, around the \ce{SO3-} groups the number of water molecules decreases and the number of hydronium ions increases. This effect is more evident for the highly hydrated films ($\lambda=22$). We also conclude that the changes between the structure of the film and the membrane increases with the hydration level.

\begin{figure*}[t]
\centering
\includegraphics[width=0.6\textwidth]{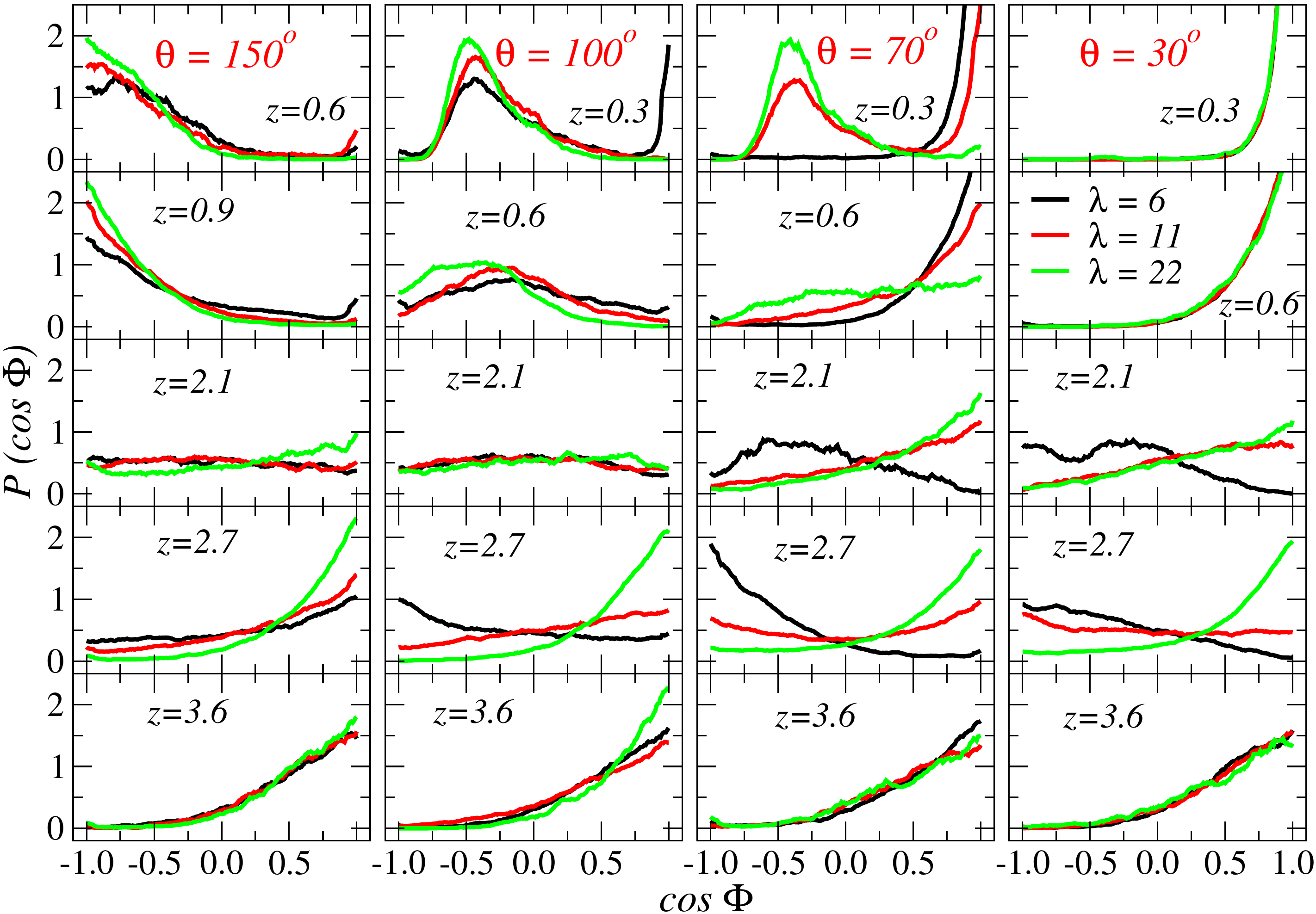}
\caption{
Probability distributions of $cos(\phi_{\ce{SO3-}})$, where $\phi_{\ce{SO3-}}$ is the angle formed by the \ce{SO3-} orientation vector $\hat{u}_{\ce{SO3-}}$ and the normal to the support, $\hat{z}$. The distributions are calculated in slabs of thickness $0.3$~nm parallel to the substrate and at the indicated distances from the support,  $z$ (in nm). In the first slab, one can observe the inversion of the \ce{SO3-} orientation when decreasing $\theta$, as discussed in the text.
}
\label{fig7:so3orient}
\end{figure*} 

\subsection{Molecular orientation profiles}
\label{subsec:orientation}
To further elucidate both global and local features of the deposited thin-films, orientational order of sulfonic acid groups in regions of the films at different distances from the support were extensively investigated. Similar information about the orientational order of water molecules has already been reported in Ref.~\cite{Borges2013}. There, we have shown that the orientation of water molecules is mainly driven by the interaction with the support, similar to the case of water molecules near Lennard-Jones smooth walls~\cite{Spohr1988,Glebov1997,Tatarkhanov2009}. 

The orientation of the \so3 groups at different distances from the support was quantified as follows: the films have been partitioned into partially overlapping slabs parallel to the support, with a thickness $\delta z=0.3$~nm. In each slab we have calculated the  probability distributions $P(cos(\phi_{\ce{SO3-}}))$, with $cos(\phi_{\ce{SO3-}})=\hat{u}_\ce{SO3-}\cdot\hat{z}$. Here, $\hat{z}$ is the unit vector normal to the support and the unit vector $\hat{u}_{\ce{SO3-}}$ is oriented normal to the plane formed by the three oxygen atoms and points toward the sulphur atom. The \so3 orientations at different distances from the support are crucial to elucidate the global ionomer orientation. As a reference, for $cos (\phi_{\ce{SO3-}}) = 1$, the three oxygen atoms face the support and lye in the $xy$-plane. 
\begin{figure}[t]
\centering  
\includegraphics[width=0.4\textwidth]{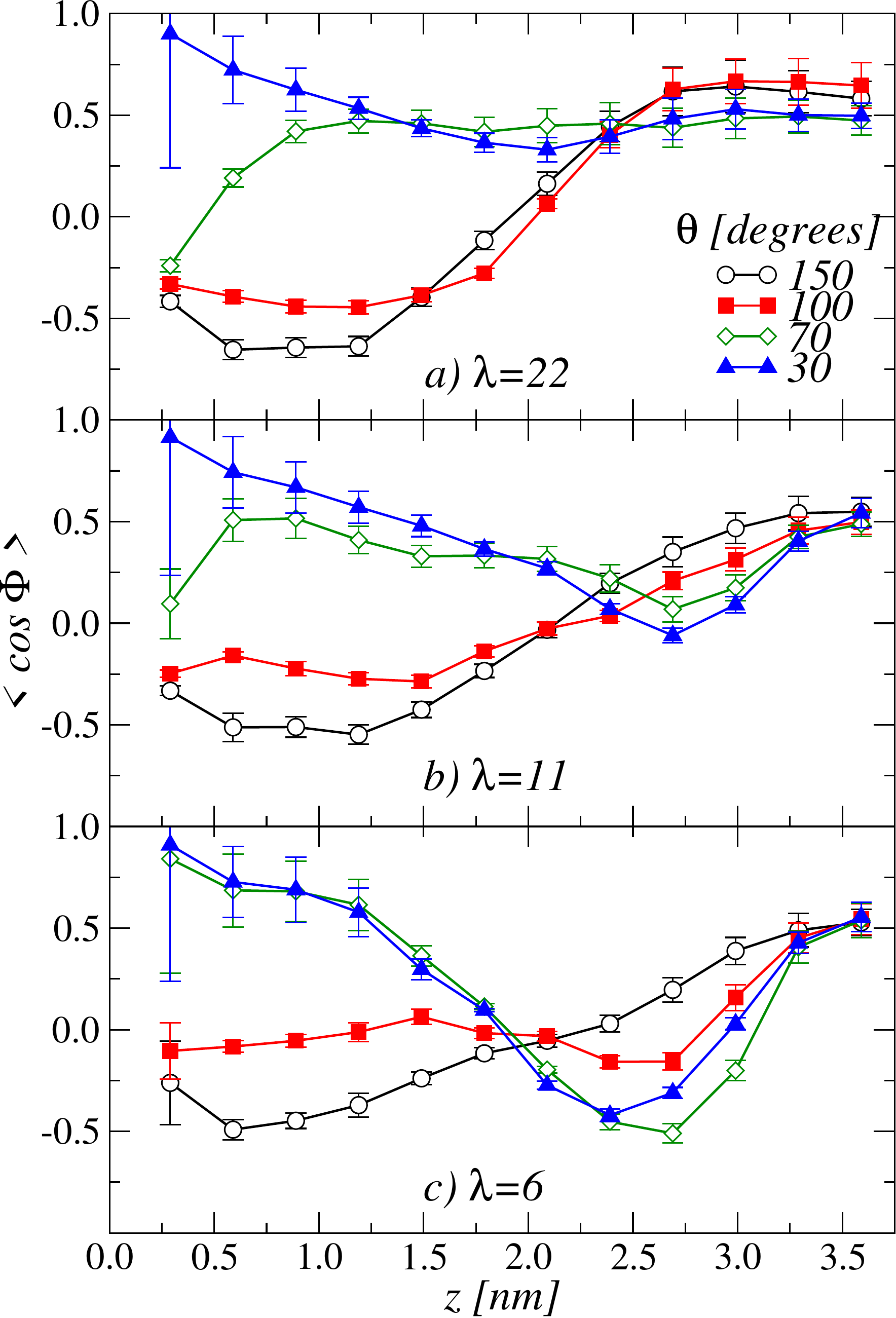}
\caption{
Average of $cos(\phi_{SO_3^-})$ as function of the distance from the surface in the films at (a) $\lambda=22$, (b) 11 and (c) 6. 
}
\label{fig8:avcos-so3} 
\end{figure}

In Fig.~\ref{fig7:so3orient} we show $P(cos(\phi_{\ce{SO3-}}))$ for the investigated films at the indicated values of $\theta$, $\lambda$ and distances from the support. Clearly, $P(cos(\phi_{\ce{SO3-}}))$ depends on the hydrophilic degree of the support. Focusing on the first layer, it is evident that in the most hydrophobic ($\theta=150^\circ$) and hydrophilic ($\theta=30^\circ$) cases, the \so3 are oriented in opposite directions. In the first case, the side chains are oriented with the sulfonate groups pointing opposite to the substrate, while in the second case, they point toward the substrate. In the intermediate cases, ($\theta=70^\circ$ and $30^\circ$), the $P(cos(\phi_{\ce{SO3-}}))$ are peaked around $-0.5$. Therefore, the three oxygen atoms point in the direction of the ionomer, with the \so3 vector forming an angle of about $60^\circ$ with the normal to the support. This orientation corresponds to side-chains aligned horizontally to the substrate. Side-chain orientational configurations parallel and orthogonal to the support are called "standing" and "lying", respectively, and have been also observed in previous simulations of the ionomer placed on top of platinum surfaces~\cite{Cheng2010,Selvan2008}.

When decreasing hydration, the degree of ionomer orientational order decreases. It is interesting to note that, in the case of $\theta=70^\circ$, the side-chains are first found in the lying position at $\lambda=22$ for gradually shifting to standing configurations, at $\lambda=6$. This indicates that water content also plays an important role in determining the side chains orientation. Indeed, in this particular low-$\lambda$ case, most part of water molecules are in contact with the substrate and,  consequently, the ionomer self-organizes to maximize the fraction of \so3 groups in direct contact with water. Details of the interface between water domains and side-chains will be further discussed below.

The data shown in Fig.~\ref{fig7:so3orient} also show that the \so3 groups are characterized by different preferential orientations in different regions within the film. In order to be more specific on this point, the evolution of the average value $\langle cos(\phi_{\ce{SO3-}}) \rangle$ across the film is illustrated in Fig.~\ref{fig8:avcos-so3}. Interestingly, side-chains orientation inversions at particular distances are evident in some conditions. This inversion is particularly clear in the cases corresponding to $\lambda=22$ (Fig.~\ref{fig8:avcos-so3}~(a)) for $\theta=150^\circ$ and $100^\circ$. Here, $\langle cos(\phi_{\ce{SO3-}})\rangle$, which is negative in the regions close to the support, steadily increases across the central region of the film eventually assuming positive values in the regions furthest from the support. Also interesting are the cases of the films at $\lambda=6$ formed on very hydrophilic supports (Fig.~\ref{fig8:avcos-so3}~(c)). For $\theta=70^\circ$ and $30^\circ$, two inversions on the side-chain average orientation are observed. Strong correlations exist in this case with the water density profiles shown in Fig.~\ref{fig4:densityProfile}~(f). Indeed, we observe the minima of $\langle cos(\phi_{\ce{SO3-}})\rangle $ at  $z\simeq$~2.25-2.75~nm, which have a significant overlap with the region where water pools have been observed ($z\simeq$~2.5-3.5~nm). This observation additionally supports the idea that side-chain orientation is mainly governed by the non-trivial distribution of water domains inside the film. An other observation originating from the data of Fig.~\ref{fig8:avcos-so3} is that at distances larger than $3$~nm, side-chain sulfonic acid groups always point toward the support, independently of the values of $\theta$ and $ \lambda$. This side-chain alignment on the top of the film is attributed in part to the ionomer/air interface. We will come back to this point in what follows.

In summary, our results demonstrate that the interaction of water molecules with the support determines the side-chains orientation. Indeed, the \ce{SO3-} groups must be embedded in water domains, to minimize the surface tension at the interface between the hydrophobic polymer backbone and water~\cite{Moore2004}. Therefore, although $\theta$ plays a mild role on the orientational properties of water molecules (as we demonstrated in ref.~\cite{Borges2013}), it has indeed a strong impact on side-chains orientation. This information is very important for the following, when we will propose a general qualitative picture for the morphology of supported Nafion thin-films. In the next Section we complete our investigation by characterizing the formation of ionic clusters across the film.
\begin{table}[b]
\centering
\begin{tabular}{ccccc}
\hline \hline 
$\lambda\backslash\theta(^\circ)$ & $150$ & $100$   & $70$    &  $30$\\
\hline \hline
22 &   1.71      &   1.87          &   3.02          &   2.96\\
11 &    3.38     &   3.10          &   3.65          &   3.33\\
6 &    5.85     &   5.64          &   4.96          &   4.99\\
\hline\hline	
\end{tabular}
\caption{
Average \so3 groups cluster sizes for the ionomer thin films at the indicated values of hydrophilicity degree $\theta$ and hydration level $\lambda$.}
\label{tb:clustersize}
\end{table}
\begin{figure}[b]
\centering  
\includegraphics[width=0.45\textwidth]{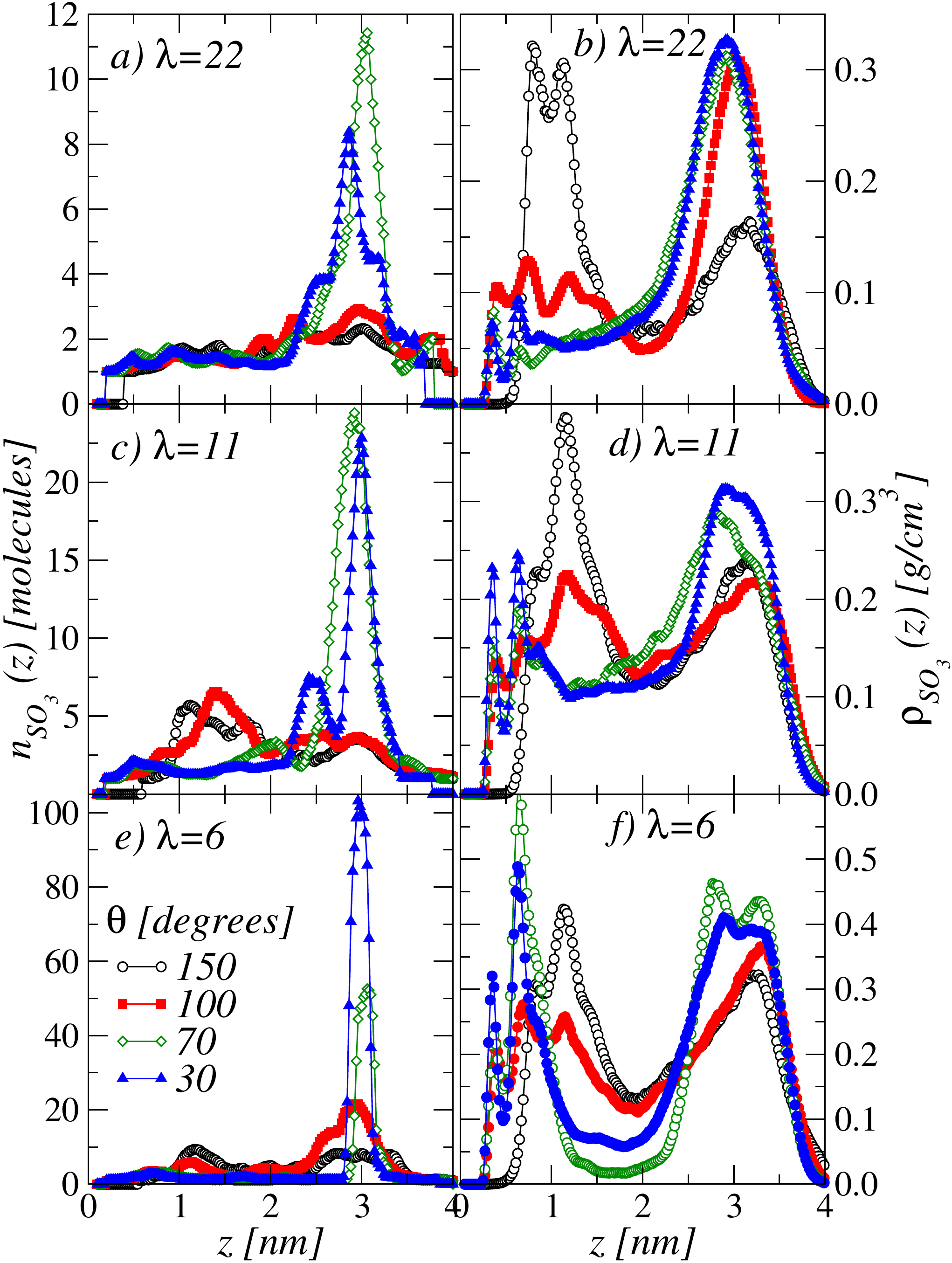}
\caption{
Average cluster size (left) and mass density distributions (right) for the \so3 groups as function of distance from the support, $z$, at the indicated values of $\lambda$ and $\theta$.}
\label{fig9:ionic-clusterZ}
\end{figure}

\subsection{Formation of ionic clusters}
\label{subsec: ionic-clusters}
Above we have shown that films present different \so3 packing features, {\em i.e.}, both coordination numbers and minimum distances between \so3 groups (Fig.~\ref{fig6:gSS-film}) change for the different investigated cases. Here we conclude our analysis by focusing on the features of ionic clusters. This information is important for proposing a general picture for the morphology of the supported films in different hydration conditions and for different wetting nature of the substrate. We have identified the ionic clusters by identifying the \so3 groups separated by a distance less than a cut-off $r_c=$~0.64~nm. The clustering analysis provide us with the probability distribution of the size of the clusters, i.e., the number of molecules pertaining to the same cluster. If a \so3 group has no nearest neighbours within the cut-off distance, it is considered as an isolated cluster of size $1$.   

In Table~\ref{tb:clustersize} we show the average cluster size for all the investigated films. At fixed $\theta$, the cluster size decreases when increasing water content, which is an expected effect due to film swelling: an increasing number of water molecules intercalates between adjacent side chains, therefore \ce{SO3-} groups form less compact agglomerates and isolated groups are found with a higher probability. The hydrophilicity degree also impacts the average cluster size in a non-trivial fashion, which possibly depends on the details of the morphology of the considered film. This result seems to be at odds with a visual inspection of the snapshots shown in Fig.~\ref{fig3:side}, where quite extended regions of condensation of  \ce{SO3-} groups are evident in particular regions of the films. To better clarify this point, we computed the average clusters size in different regions of the film, as a function of the distance $z$ from the substrate. In Fig.~\ref{fig9:ionic-clusterZ} we plot the average cluster sizes $\langle S_{\ce{SO3-}}(z)\rangle$ (left), together with the sulfonic acid mass density distributions $\rho_{\ce{SO3-}}(z)$ (right). This helps us in underlining the regions where the presence of  \ce{SO3-} groups is relevant. For all values of $\lambda$, at $\theta=30^\circ$ and $70^\circ$, the $\langle S_{\ce{SO3-}}(z)\rangle$ curves clearly indicate the formation of very extended clusters at distances larger than $2$~nm from the support, in the top part of the film, closer to the ionomer/air interface. This is consistent with the high \so3 mass density in this region. However, we also note that, for the cases $\theta=150^\circ$ and $100^\circ$, the distribution of average cluster sizes does not show any pronounced peak, despite the presence of well defined maxima in the $\rho_{\ce{SO3-}}$ curves. In conclusion, the formation of \so3 clusters seems not to be simply determined by the distribution of \so3 but is apparently controlled by the details of the morphology of the film. Also, we emphasize that ionic clustering should play a crucial role on water dynamics. In general, \so3 group cluster has a strong impact on hydrogen binding between side-chains, and determines both water binding and the different mechanisms of proton transport~\cite{Kreuer2000,Elliott2007}. 

\subsection{Water clusters and connectivity}
\label{sec: water-domains}
\begin{figure}[]
\centering
\includegraphics[width=0.38\textwidth]{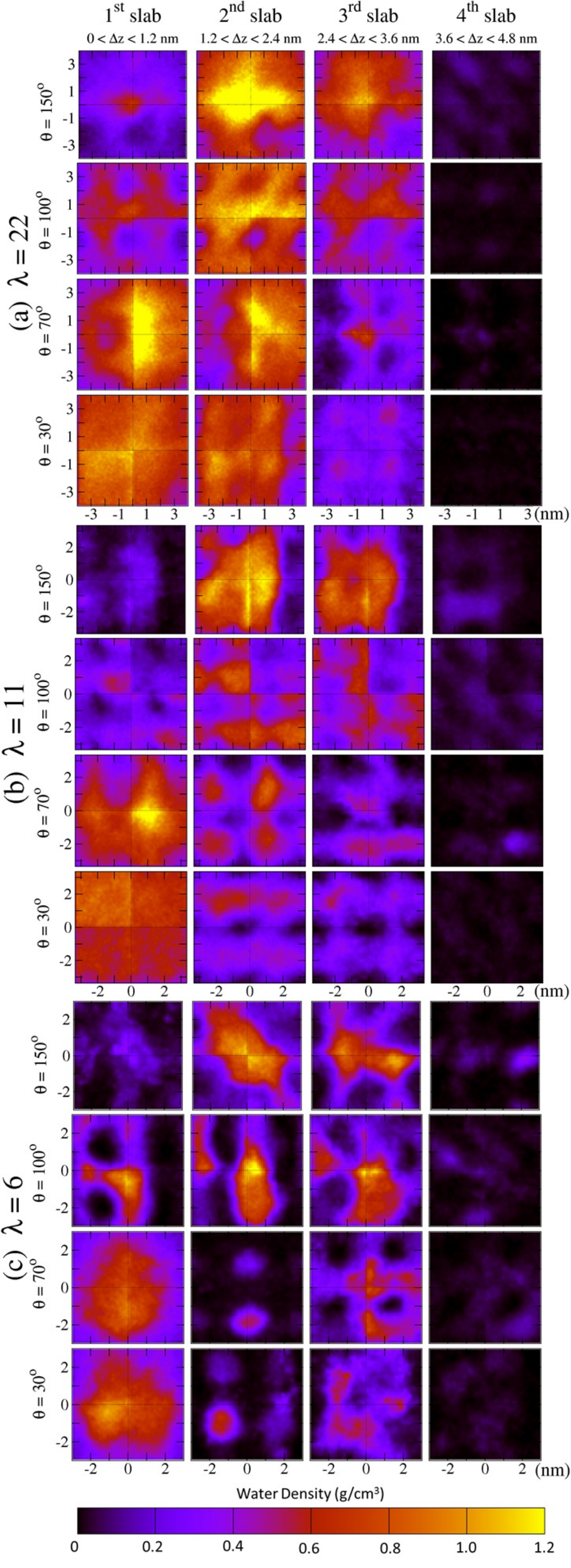}
\caption{
Contour plots of water density for $\lambda=$ 22 (a) 11 (b) and 6 (c), calculated in slabs at the different indicated distances from the substrate.
}
\label{fig10:map-water}
\end{figure}

We now focus on the topology of the domains formed by the water molecules, and investigate both shape and connectivity of the hydrated domains. We have characterized the water mass density distributions in planes parallel to the substrate, by partitioning the film in four slabs of thickness $1.2$~nm and computing the projected water density distributions on the $xy$-plane, averaged over the trajectory. Our data are plot in the form of color maps in Fig.~\ref{fig10:map-water}. Here a lighter color (yellow) identifies regions where water density is higher, while darker color characterizes regions where the presence of ionomer is significant.

We first consider the maps in Fig.~\ref{fig10:map-water} for the most hydrophobic cases ($\theta=150^\circ$). Water is concentrated in the second and third slabs, and at $\lambda=22$, a quite homogeneous distribution suggests that water molecules form a unique layer parallel to the support and confined by two ionomer layers separated by a distance of about $\sim 2.4$~nm. The side-chains pertaining to the facing ionomer layers point toward the water layer, with Nafion chains adopting a "sandwich" morphology. In contrast, when decreasing water content, the water pool tends to be concentrated in the central region of the film, surrounded by the ionomer. This is particularly evident for $\lambda=6$, where water molecules form an elongated domain and seems to suggest an inverted micelles morphology, with ellipsoidal or cylindrical micelles shape oriented parallel to the substrate. In the intermediate case, $\theta=100^\circ$, although we do not observe any percolating water-rich region that could be considered as a continuous water layer, water can still form extended agglomerates in the three slabs closer to the wall. For $\lambda=6$, these water "pools" are well delimited and seem to be connected in adjacent slabs. We can also observe a few ionomer "barriers" (indicated by the darker color in the middle of the maps) connecting hydrophobic domains in adjacent slabs.  At high hydration, $\lambda=22$, the formation of "pools" is less clear, water being quite homogeneously distributed in all regions, with  the ionomer well hydrated everywhere. 

In the most hydrophilic cases, $\theta=30^\circ$ and $70^\circ$, water distributions are similar, and the largest water domain forms in contact with the substrate, as expected. For $\lambda=22$, the amount of water is also significant in the second slab. This suggests that water forms a thick continuous layer between the substrate and the ionomer which accumulate on the top of the film, at the interface with air. As a result, these films adopt a completely phase-separated bi-layer configuration. When $\lambda$ decreases, water domains become less homogeneous already beyond the first considered layer, and the formation of disconnected pools in the middle of the film is observed. For $\lambda=6$, water is mostly concentrated in the first and third slab, suggesting a morphology with alternated water-poor and water-rich layers. Also, a single narrow water channel forms, directly connecting the two otherwise disconnected water domains. We finally observe that in all cases the fourth furthest slab is not populated by water molecules, consistent with a hydrophobic interface with air, mostly composed by the ionomer backbones with the side-chains pointing toward the substrate~\cite{Bass2010}.
\begin{figure}[t]
\centering
\includegraphics[width=0.49\textwidth]{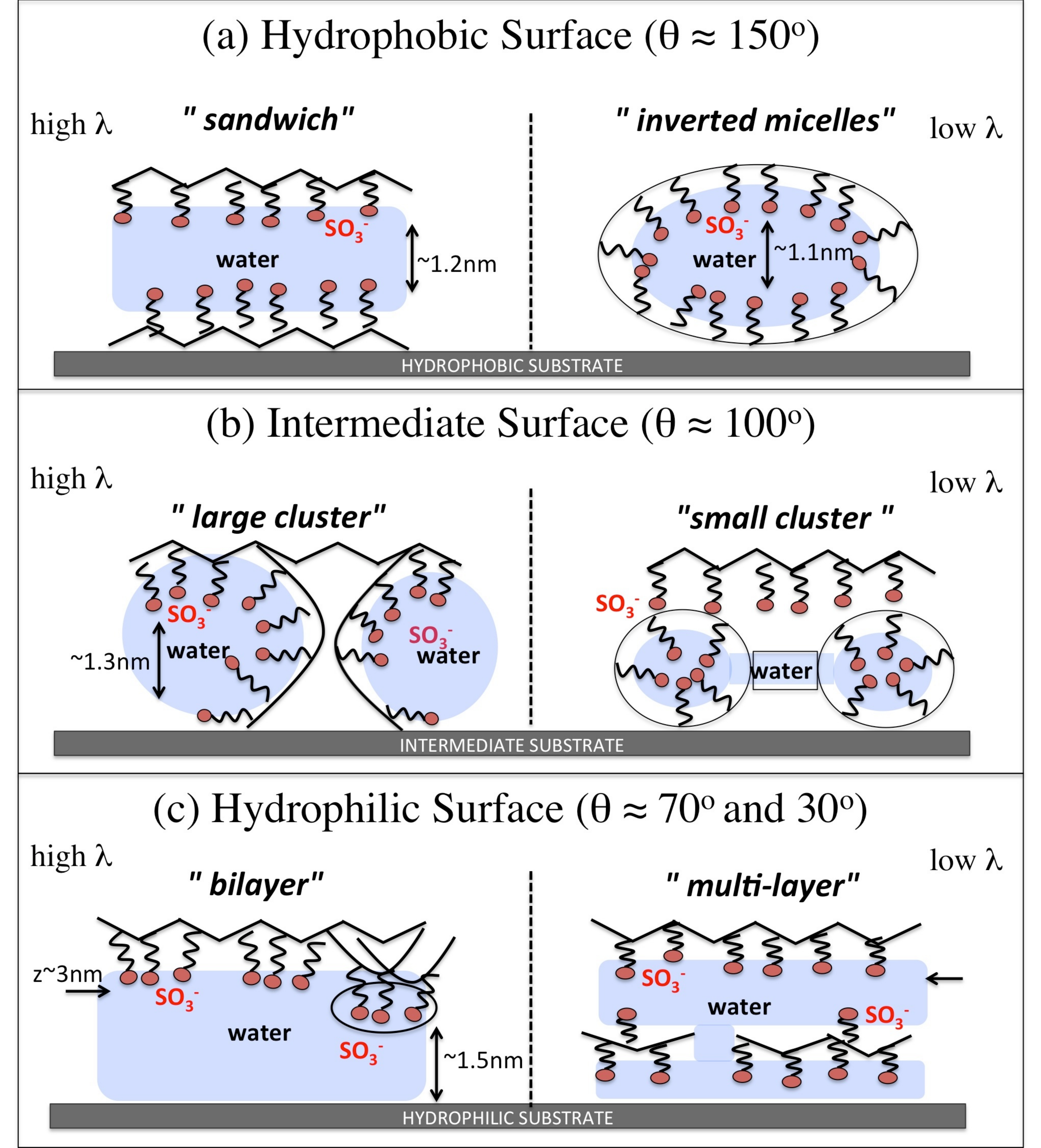}
\caption{
Qualitative picture of film morphologies, at different values of $\theta$, ranging from highly hydrophobic (top) to very hydrophylic (bottom) and different hydration levels $\lambda$ (high and low hydration on left and right, respectively).
} 	
\label{fig11:scheme-film}
\end{figure}

\subsection{A qualitative picture for morphology}
\label{subsec:general-morphology}
Based on the analysis presented in the previous Sections we are now in the position to draw a general picture of the morphology of the supported hydrated Nafion thin films, at different hydration levels and for for varying wetting nature of the support. Despite the qualitative nature of our conclusions, this is the most important message of the present work. We schematically represent the expected morphology of the thin-films in the different conditions as cartoons in Fig.~\ref{fig11:scheme-film}. The \so3 groups are represented by red beads, side chains by spring-like symbols and polymer backbones by solid black lines. Water pools are the blue domains. In summary, with reference to the wetting character of the support, we classify the typical morphologies in three classes: 

{\bf 1.~Hydrophobic} The film at high hydration (left) shows a typical "sandwich" structure, constituted by a sequence of layers of different nature (Fig.~\ref{fig11:scheme-film}~(a)). This is in agreement with the experimental observations of Refs.~\cite{Dura2009,Wood2009}. Nafion backbones are therefore in direct contact with the substrate, with the sulfonic acid groups pointing upward, toward the water domain. On the top of the water pool, a reversed structure sulfonic groups/polymer backbone is observed, with a completely hydrophobic film/air interface.  At low water content (right), the ionomer folds around the water domain, forming an inverted-micelle structure, which reminds the experimental observations of Refs.~\cite{Bass2010, Bass2011}. More precisely, in our simulations the ionomer folds into a inverted-micelle cylinders of diameter $\simeq 4$~nm and with the symmetry axis parallel to the support, as one can observe in the water maps in Figure~\ref{fig10:map-water} 

{\bf 2.~Intermediate} In this case the ionomer film organizes into a configuration with interconnected water "pools" (Fig~\ref{fig11:scheme-film}~(b)). The film/substrate interface is characterized by both the presence of ionomer and water, while the film/air interface still has a hydrophobic character, with the side-chains of the ionomer pointing toward the substrate. Hydration level mostly impacts the size of water pools, which decreases by decreasing $\lambda$. In general, the local structure of the film in this case is very similar to the case of the membrane and no evident phase separation parallel to the support is present.

{\bf 3.~Hydrophilic} Thin films in contact with very hydrophilic substrates are organized in well-separated water and ionomer layers (Fig~\ref{fig11:scheme-film}~(c)). In high hydration conditions (left ) water floods the substrate and the ionomer accumulate at the top, with the hydrophobic polymer backbone in contact with air. For lower values of $\lambda$ (right), the ionomer approaches the support. This behavior is not driven by a direct interaction with the substrate, but rather indirectly due to  the interaction of the side chains with the water layer in contact with the support. In this case the film can adopt a multilamellar configuration with multiple water layers parallel to the substrate and separated by ionomer domains. Adjacent water layers can be locally connected by water channels, which form dynamically but seem to be quite stable. This picture originating from our data is also consistent with the experimental observations of Refs.~\cite{Dura2009,Wood2009}, where the Authors discovered lamellar structures, formed close to hydrophilic substrates and composed of alternating water-rich and Nafion-rich thin layers. 

We conclude this Section by observing that in this work we have considered very thin films of about $4.5$~nm and therefore showed that the wetting nature of the support strongly impacts morphology on length scales of the order of a few nanometers. However, we have also underlined that our qualitative picture seems to be in agreement with experimental observations on films of much larger thickness. We therefore conjecture that the structure of real films could be the results of a geometrical tiling, where the local building blocks are morphologies similar to the ones of Fig.~\ref{fig11:scheme-film}. How this tiling extends from the substrate to the ionomer/air interface in real systems is an open issue. In what follows we will discuss how the qualitative features summarized above can be relevant for PEMFC technology. 

\section{Nafion thin-films morphology and PEMFC technology}
\label{sec:PEMFC-technology}
In this Section we discuss the relevance of our findings in the understanding of the catalyst layer features, a crucial issue in the PEMFC technology. From our analysis, the ionomer morphology is expected to impact the catalyst layer activity as follows. First, a strong effect can be envisaged on the transport features of water and hydronium complexes close to the catalyst and the catalyst/support interfaces. Indeed, we have shown in our previous publication~\cite{Borges2013} that complex morphology changes can result in a highly heterogeneous transport behaviour of water across the film. In particular, the extent of the heterogeneity seems to be directly controlled by the wetting character of the substrate and increases steadily by increasing the hydrophilicity character of the support~\cite{Borges2013}. 

Second, our findings could also be relevant for a better understanding of the ionomer/catalyst interface. This is the region where the electrochemical reactions governing a PEMFC operation take place. In the actual device, two phenomena directly affect the reaction kinetics: adsorption of chemical species and the formation of the electrochemical double layer. Detailed descriptions of these mechanisms are not possible with our level of description, which cannot account for electrochemical activity. We can however speculate about the impact of the ionomer structural organization on these phenomena. 

Third,  analysis of the (top) film/air interface is relevant to understand its impact on the water and reactant gases transport inside catalyst layer pores (in the CL gas phase). The upper surface of the film plays an important role in the hydrophilicity of the catalyst layer pores, which in turn impacts water management during operation conditions. Moreover, the reactant gases in the gas phase (\eg \ce{O2} and \ce{H2}) must cross the film in order to reach the catalyst surface where the reactions take place. Below we will describe the ionomer/air interface and its possible impact on the water and gases absorption and water management. 

In what follows we explore these points in details, by characterizing the interfacial regions, \ie immediately adjacent to the substrate and at the top of the  film. We will first analyse ionomer adsorption and overall substrate coverage for different wetting nature of the support. Next, we will investigate the main features of the charge distribution close to the substrate. Finally, we will characterize the ionomer/air  interface.
\begin{figure}[t]
\centering  
\includegraphics[width=0.7\textwidth,angle=-90,origin=c]{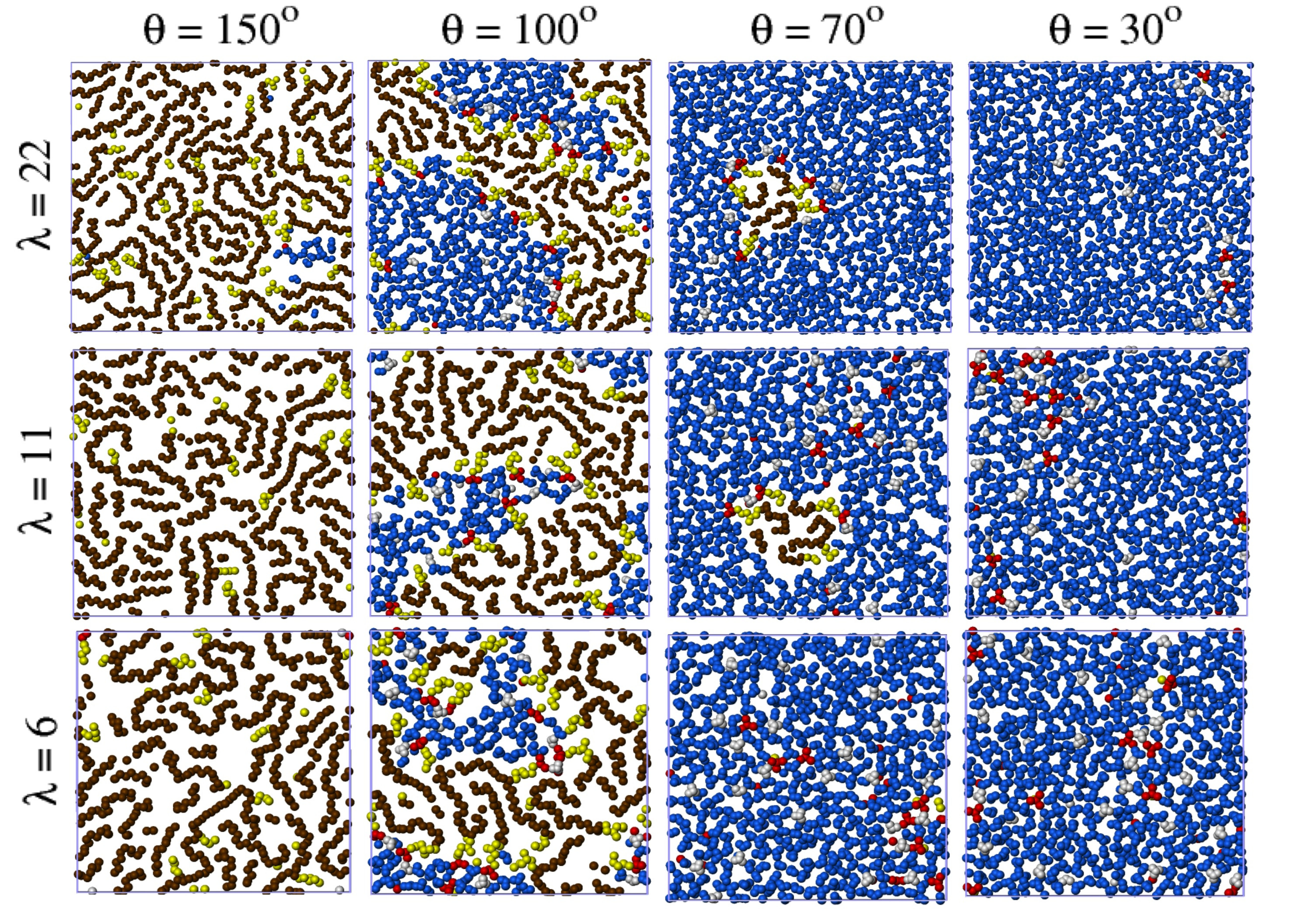}
\caption{
Snapshots of the adsorption region, which extends to $z\simeq 0.56$~nm from the support. The backbone segments beads are plotted in brown, side-chains hydrophobic segments in yellow, the \ce{SO3-} groups in red, water molecules in blue and hydronium complexes in white. 
}
\label{fig12:snapshot-adsorptionRegion}
\end{figure} 

\subsection{Ionomer adsorption}
\label{subsec:film-wall}
In the CL, the catalyst (Pt and/or Pt-alloy) surfaces can react with water, hydronium ions or other chemical species~\cite{Subbaraman2010}. Although in this work electrochemical reactivity of the substrate is not accounted for, we are in the position to characterize the overall surface coverage. This should depend on the details of the ionomer distribution immediately adjacent to the substrate, which corresponds to the first peak in the mass density profiles of Fig.~\ref{fig4:densityProfile}. In Fig.~\ref{fig12:snapshot-adsorptionRegion} we show typical snapshots of the adsorption region, which extends to $z\simeq 0.56$~nm from the support. In the case of hydrophobic substrates, $\theta=150^\circ$, and at any degree of hydration, the ionomer is adsorbed via the backbone, as also observed in simulations of an ionomer adsorbed on graphitized carbon sheets~\cite{Mashio2010}. For the case of intermediate hydrophilicity, $\theta=100^\circ$, a more balanced presence of water, backbone segments and side-chains is observed. In the most hydrophilic cases, $\theta=70^\circ$ and $30^\circ$, limited adsorption of the ionomer is still observed, which takes place via the sulfonate groups (red beads in Fig.~\ref{fig12:snapshot-adsorptionRegion}). 

\begin{figure}[t]
\centering  
\includegraphics[width=0.4\textwidth]{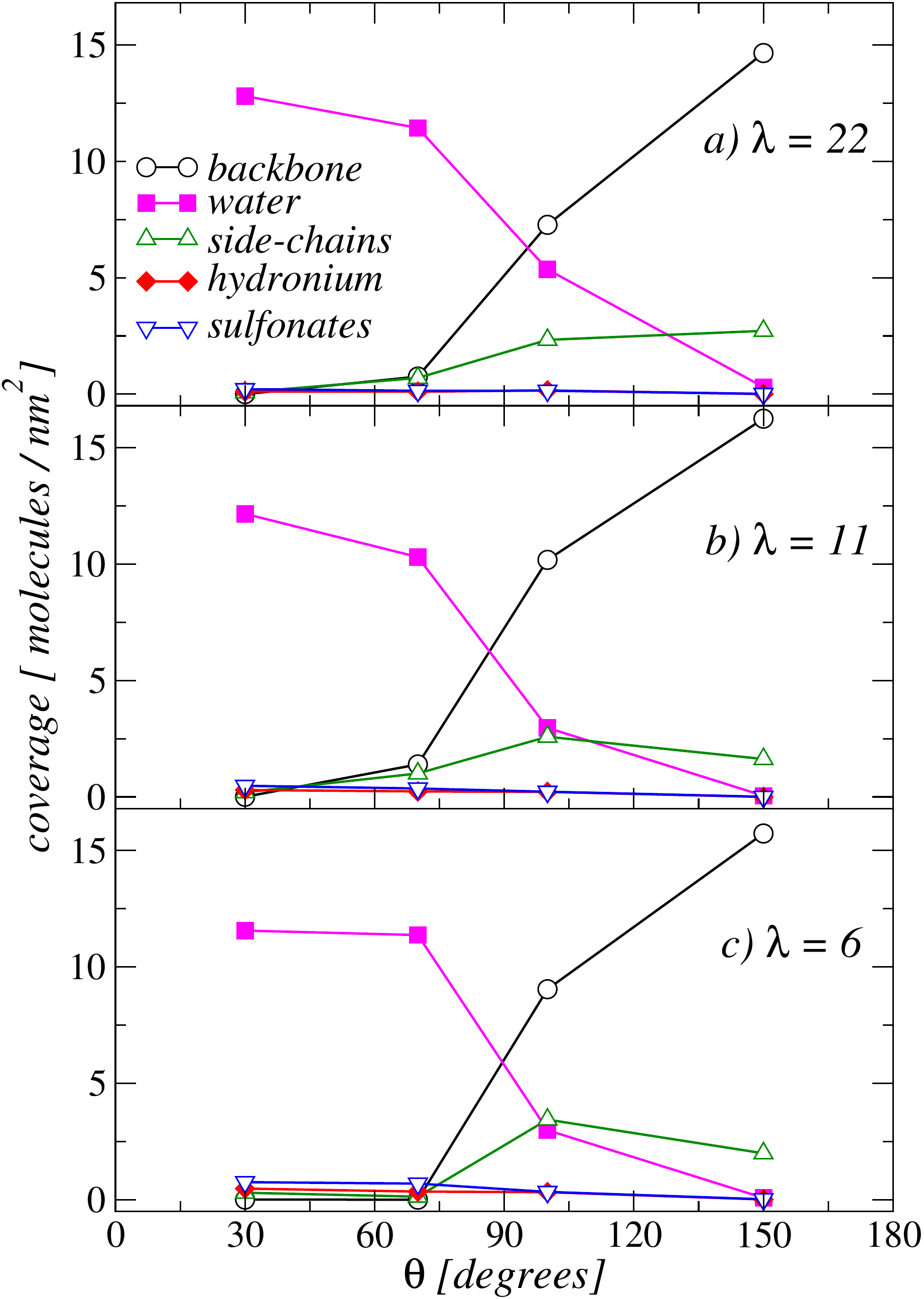}
\caption{
Ionomer backbone, water, side-chains, \so3 and hydronium complexes coverage as a function of the substrate contact angle, at the indicated values of the hydration levels.
}
\label{fig13:coverage}
\end{figure}

The average substrate coverage for the ionomer backbones, \water, side-chains,  \hy and \so3 groups are shown in Fig.~\ref{fig13:coverage} for all thin-films investigated. The coverage is defined here as the number of molecules within the adsorption region per unit of area. The data in Fig.~\ref{fig13:coverage} clearly show an inversion of surface coverage following the hydrophilicity degree of the support. In contrast, water content does not seem to significantly modify ionomer backbone or water coverages. Indeed, by decreasing water content from $\lambda=22$ to $\lambda=6$, backbone coverage changes from $14.65$ to $15.72$~molecules/nm$^2$ for the most hydrophobic case, while water coverage reduces from $12.80$ to $11.56$~molecules/nm$^2$ for the most hydrophilic case. The reduction of water coverage is compensated by the increases of \hy and \so3 coverages. \so3 coverage increases from $0.003$ to $0.007$~molecules/nm$^2$ while the \hy coverage changes from $0.008$ to $0.015$~molecules/nm$^2$. Hence, the number of adsorbed \so3 groups is higher for $\lambda=$~6 and 11, and they are well dispersed on the surface. In contrast, for $\lambda=22$, the \so3 groups can be found in more agglomerated configurations. Overall, Figs.~\ref{fig12:snapshot-adsorptionRegion} and~\ref{fig13:coverage} further corroborate our previous observation of a transition from a predominant backbone coverage to predominant water coverage, when increasing the hydrophilic character of the substrate. However, even for most hydrophilic cases adsorption of the ionomer is still observed and occurs mainly via \so3 groups. The adsorption of \so3 is more evident when the hydration of the film is lower.

During PEMFC operation, oxidation and reduction reactions occurring on the top of catalyst surfaces strongly depend on surface coverage of reactants and spectator species~\cite{Franco2006,Malek2011a,DeMorais2011}. Our results shows that water molecules and hydronium ions can be found away from the catalyst surface, in the case where the wetting nature of the substrate is not favourable. The adsorption of the ionomer could block the adsorption of reactant species, reducing the area where the electrochemical reaction occurs. Note that this behaviour is usually overlooked when addressing the issue of increasing Pt utilization in PEMFCs.

Also important for PEMFC development is to clarify the impact of ionomer adsorption in ORR mechanism. It is well know that the kinetics of the ORR is sensitive to the nature of adsorption of spectator species~\cite{Markovic2002}. For example, specific adsorption of sulfonate anions has an important deactivation effect on the ORR. The extent of this feature correlates with the strength of the catalyst-sulfonate bond (the strenght of \so3 adsorption)~\cite{Subbaraman2010, Subbaraman2010a}. Various factors can influence the chemical nature of \so3 adsorption, including nature of the counter-cation, extent of \so3 agglomeration within the ionomer, length and spacing between side chains adjacent along the backbone. Our results show that the \so3 groups are adsorbed in different configurations, \eg, both clustered and dispersed. This should affect the chemical nature of the \so3 adsorption, and ultimately affect the electrochemical potential that drives the electrochemical reactions. 

To conclude this Section, we observe that cell reactions are also governed by the structural properties of the Electrical Double Layer (EDL) formed close to the electrode surface~\cite{Quiroga2014}. Unfortunately, standard electrochemical theories normally used to describe the EDL, completely ignore the heterogeneous environment created by the adsorbed ionomer, which affects both charge and potential distributions~\cite{WangLRoudgarA2009, Krapf2006, Zhdanov2006561, Zhdanov2004, Zhdanov2008, Biesheuvel2009}. In contrast, our findings clearly show that the ionomer dictates the distribution of charges very close to the surface (as indicated by the ionic distributions shown in Fig.~\ref{fig9:ionic-clusterZ}) and, as a consequence, the over-potential at the reactant-electrode distance ($\sim$~0.2-0.5~nm) is also affected. Moreover, considering the different ionomer morphologies that may be found inside the CL, it is not much to say that the reaction rates are far from being uniformly distributed inside CL. Our results also strongly support the existence of a non-uniform spatial distribution of reaction rates, due to the complexity of the ionomer structure. An effective control of the ionomer morphology could therefore provide a valuable path for further development of PEMFC technology, for optimizing electrochemical interface and reducing ionomer inhibition.

\subsection{The ionomer/vacuum interface}
\label{subsec:film-air}
The morphology of the Nafion/vacuum interface has recently received special attention, also due to its importance in ionomer water uptake~\cite{Bass2011}. This interface includes the hydrophobic ionomer backbones which are exposed to the gas phase, and the underneath hydrophilic side-chains, pointing toward the water-rich domains. It is considered responsible for the so-called Schroeder's paradox, {\em i.e.,} a different Nafion water uptake from a liquid solvent or its vapour~\cite{Freger2009, Choi2003}. 

\begin{figure}[t]
\centering  
\includegraphics[width=0.7\textwidth,angle=-90,origin=c]{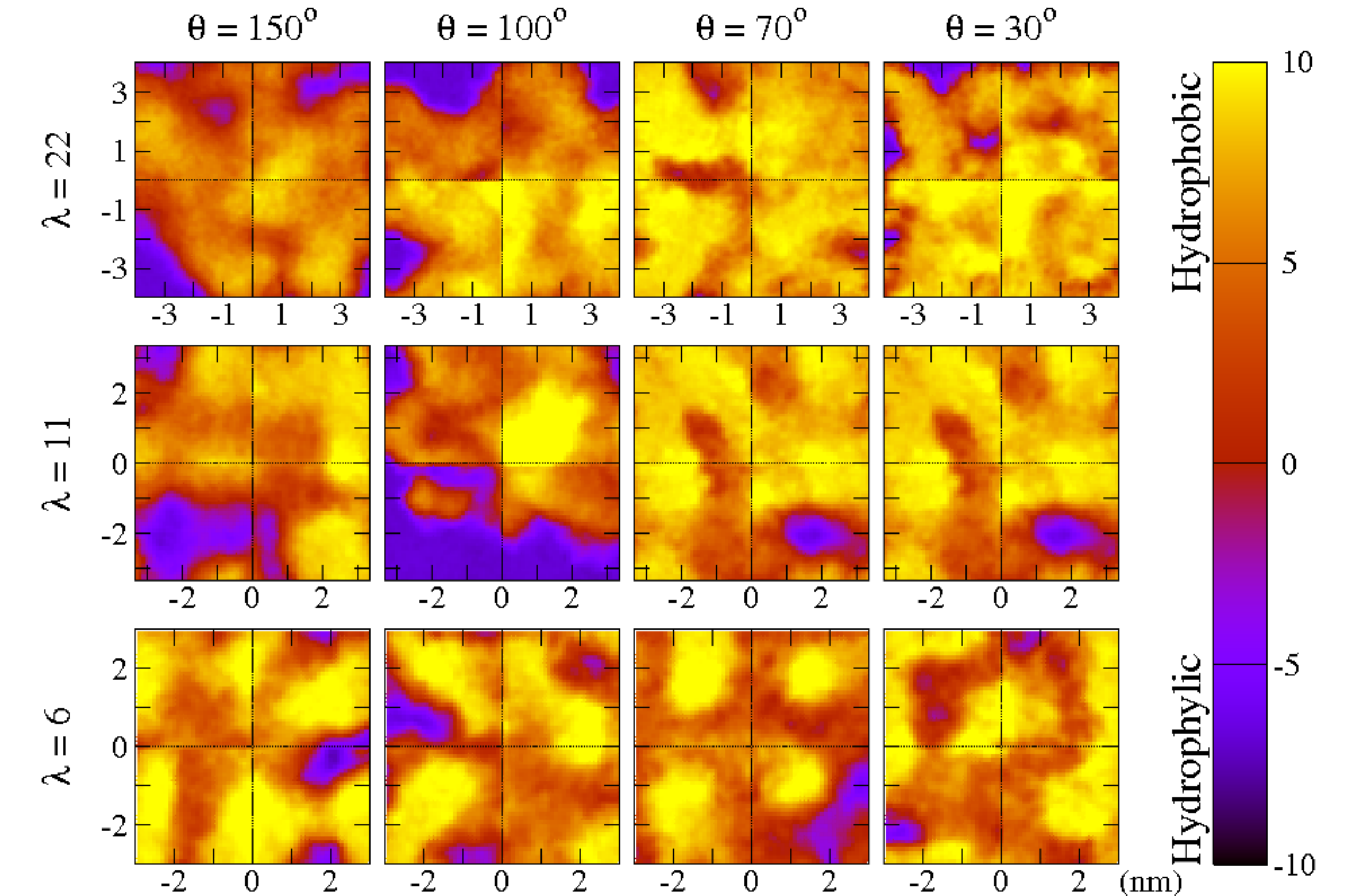}
\caption{
Colour maps of the wetting character of Nafion thin-film ionomer/vacuum interface. Hydrophobic and hydrophilic regions are in yellow and blue, respectively. The technique used for determining the maps is described in details in the text.
}
\label{fig:filmsurfacemap}
\end{figure} 
\begin{figure*}[]
\centering  
\includegraphics[width=0.9\textwidth]{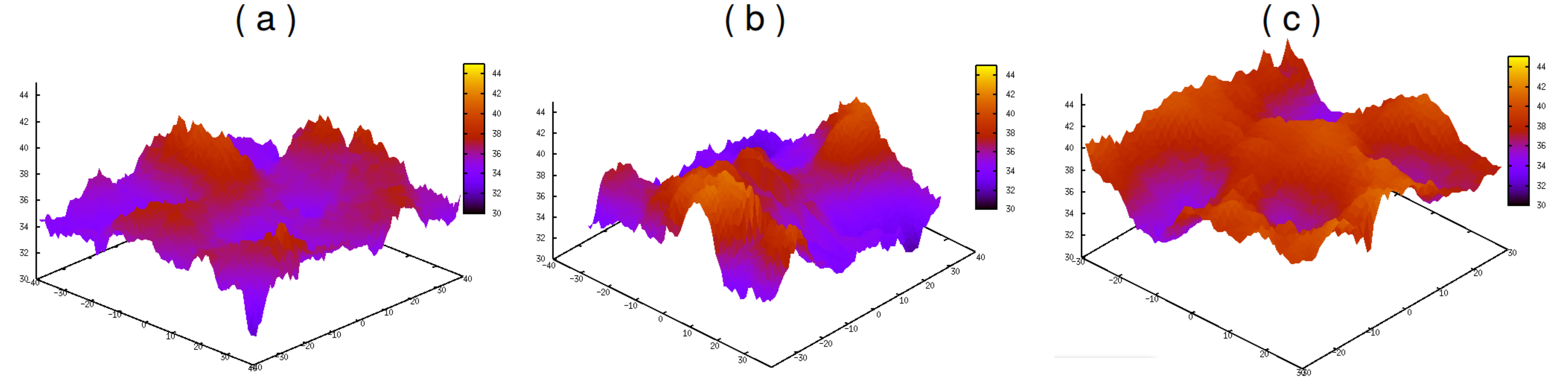}
\caption{
$xy$-contour maps of the $z$-position of atoms at the ionomer/vacuum interface for thin-films at $\theta=70^\circ$ and hydration levels $\lambda=$ 22 (a), 11 (b), and 6 (c).}
\label{fig:roughnessmap}
\end{figure*}

In order to explore the wetting nature of the ionomer/vacuum interface, we have determined spatial color maps of the local hydrophilic/hydrophobic character of the interface. In our calculations we have considered the atoms pertaining to polymer backbones and side-chains (different than sulfonate groups) as hydrophobic, while hydrophilic species included sulfonate groups, water molecules and hydronium ions. We have identified the ionomer/vacuum interface as the region with $3.0\le z\le 4.5$~nm. This region was partitioned in a regular grid, with cubic cells of volume $0.2\times0.2\times1.5$~nm$^3$, for all considered cases. We next attributed to each cell $i$ the difference in volume associated to hydrophobic and hydrophilic atoms in the cell, $\delta V^i = V^i_{phobic}-V^i_{philic}$. A negative value of $\delta V^i$ therefore corresponds to a mostly hydrophilic cell, a positive value to a hydrophobic one. The volume associated to each atom was computed by considering the value of the corresponding Lennard-Jones interaction parameter $\sigma$ as the effective diameter of the atom. We considered an average over an ensemble of $10^3$ configurations for each film.

In Fig.~\ref{fig:filmsurfacemap} we show the wetting maps for all films considered. The color range interpolates from strongly hydrophobic (yellow) to very hydrophilic (violet) regions. Thin films clearly present a compact and extended hydrophobic layer on the top in all cases, as already demonstrated above. However, violet regions are evident for $\theta=150^\circ$ and $100^\circ$ at high values of $\lambda$, which result from significant number of water molecules which accumulate immediately below the polymer backbone. In contrast, films with $\theta=70^\circ$ and $30^\circ$ present hydrated regions of very limited extent. These data suggest that the hydrophobicity of the ionomer/vacuum interface is particularly pronounced in the case of films formed on very hydrophilic substrates. At the lowest water contents, the films present similar surface hydrophobicity at all $\theta$ values. 

Our results also suggest that tuning the film/substrate interaction can modify the Nafion ionomer/vacuum interface morphology. For instance, the substrate with $\theta=30^\circ$ determines an interface configuration where the entire water content is confined under the polymer, whereas the ionomer backbone forms a "crusty" hydrophobic layer. This "crust" should present high resistance to deformation, which could decrease water uptake and lead to transport losses during PEMFC operation. It could also prevent reactants (\ce{O2} and \ce{H2} coming from the CL pores) to cross the thin-film for reaching the catalyst sites. In contrast, the films formed on the substrate with $\theta=150^\circ$, are characterized by a configuration where a fraction of the ionomer backbone is in direct contact with the substrate. This reduces the concentration of polymer backbone at the interface with vacuum and, as a consequence, increases the presence of water. Clearly, this interface should be more favourable for water absorption, which is in contrast  with the results of Ref.~\cite{Modestino2012} where, however, thin-films about $20$ times thicker than the ones considered here were investigated.
\begin{table}[b]
\centering
\begin{tabular}{cccc}
\hline \hline 
$\theta(^\circ)\backslash\lambda$& $22$ & $11$ & $6$\\
\hline \hline
$150^\circ{}$     &   0.16    &   0.25     &   0.29 \\
$100^\circ{}$ 	  &   0.21    &   0.56     &   0.29 \\
$70^\circ{}$  	  &   0.13    &   0.46     &   0.24 \\
$30^\circ{}$  	  &   0.25    &   0.44     &   0.30  \\
\hline\hline	
\end{tabular}
\caption{
Roughness coefficient $R$~(nm) for the ionomer/vacuum interface, calculated as discussed in the text.
}
\label{tb:roughness}
\end{table}

The hydrophobic "crusty" ionomer/vacuum interface is characterized by a certain degree of roughness, which depends on the hydration conditions. Roughness can be quantified as the vertical deviation of the real surface compared to its ideal form, defined as the average vertical position of the interface. We can thus define a mean-squared roughness coefficient as $R^2=1/N \sum\limits_{i=1}^{N}(Z_i-\bar{Z})^2$, where $Z_i$ denotes the $z$-coordinate of the exposed atom $i$ at the interface, $\bar{Z}$ is the average $z$-position of the surface atoms, and $N$ is the number of the surface atoms~\cite{Huang2012}. Surface atoms are identified as those with no other atoms in a square prism of edge $0.1$~nm and height $5$~nm above them. 

In Fig.~\ref{fig:roughnessmap} we show the $xy$-contour maps of the $z$-position of atoms at the ionomer/vacuum interface, for the case $\theta=70^\circ$, at the indicated values of $\lambda$. Table~\ref{tb:roughness} reports the values of $R$ for all films studied. The roughness of the films surface assumes values in the range $0.13\div 0.56$~nm, which can be compared to an experimental value of the roughness of Nafion films in contact with air of $0.35$~nm~\cite{Bass2010}. Interestingly, the roughness of the films at intermediate hydrophilicity, $\theta=100^\circ$, are slightly higher when compared to other films. This can be attributed to the disordered cluster configurations described above. According to Bass {\em et al.}~\cite{Bass2010}, the morphology of these interfaces is stable as long as the water vapour is not saturated. At that point, the hydrophobic layer should deform and the buried hydrophilic groups eventually migrate to the surface. However, when the surface is initially hydrophobic (especially at low water contents), the high energetic and kinetic barriers associated with the rearrangement of many chemical groups, may keep the ionomer kinetically trapped in this state for very long times~\cite{Bass2010}.

\section{Conclusions and Perspectives}
\label{sec:conclusions}
We have studied by Molecular Dynamics simulations the formation of Nafion ultra-thin films in contact with unstructured flat supports, characterized by their global wetting properties only. By tuning a single control parameter, $\epsilon_{w}$, we have been able to investigate in an unique framework an extended range of environments peculiar of the PEMFC catalyst layer, ranging from strongly hydrophobic (carbon-like) to very hydrophilic (platinum-like). The hydrophilicity degree of the substrate was estimated by computing the contact angle of a water droplet gently deposited on it. We considered four substrates, from strongly hydrophobic, through intermediate and  mildly hydrophilic to very hydrophilic. Also, three hydration levels were considered, in order to investigate the role played by water content. Self-assembled instances of the thin-films corresponding to these very diverse conditions were analysed in details, in terms of their structural properties. Based on a very extended data sets, we have been able to propose a general picture for Nafion supported thin films morphology for variable wetting nature of the substrate and hydration conditions.

Our data show that variations in the hydrophilic character of the substrate have strong impact on film morphology. This ranges from a sandwich structure, where an extended water pool is sandwiched by ionomer sheets, to a bilayer configuration. In this case water floods the interface with the substrate and polymer mostly accumulate at the top, at the interface with air. By decreasing water content, films convert into inverted micelles and multilamellar structures, for hydrophobic and hydrophilic supports, respectively.  We have also discovered that, in contrast to the sandwich structure, the bilayer structure shows large and compact \so3 agglomerates, resulting in a poor hydration of \hy and \so3. Analysis of surface coverage showed a clear transition from predominant backbone coverage to predominant water coverage, when switching from hydrophobic to hydrophilic surfaces. Finally, we have shown that tuning the hydrophilicity of the substrate it is possible to modify the film/vapour interface.

The results presented in this work could be of interest for optimization of the catalyst layer performances and further development of PEMFC technology. We have shown that it is indeed possible to control the main morphological features of the films by tuning the wetting nature of the substrate. Therefore, the use of appropriate substrates could be highly attractive for controlling some aspects such as ionomer coverage, proton accessibility to the active surface, \so3 adsorption, among others. This would optimize the electrode/electrolyte interface, in order to create electrochemical environment favourable to enhance cell reaction rates.
\bibliographystyle{achemso}
\bibliography{references}

\providecommand{\latin}[1]{#1}
\providecommand*\mcitethebibliography{\thebibliography}
\csname @ifundefined\endcsname{endmcitethebibliography}
  {\let\endmcitethebibliography\endthebibliography}{}
\begin{mcitethebibliography}{100}
\providecommand*\natexlab[1]{#1}
\providecommand*\mciteSetBstSublistMode[1]{}
\providecommand*\mciteSetBstMaxWidthForm[2]{}
\providecommand*\mciteBstWouldAddEndPuncttrue
  {\def\EndOfBibitem{\unskip.}}
\providecommand*\mciteBstWouldAddEndPunctfalse
  {\let\EndOfBibitem\relax}
\providecommand*\mciteSetBstMidEndSepPunct[3]{}
\providecommand*\mciteSetBstSublistLabelBeginEnd[3]{}
\providecommand*\EndOfBibitem{}
\mciteSetBstSublistMode{f}
\mciteSetBstMaxWidthForm{subitem}{(\alph{mcitesubitemcount})}
\mciteSetBstSublistLabelBeginEnd
  {\mcitemaxwidthsubitemform\space}
  {\relax}
  {\relax}

\bibitem[Eikerling \latin{et~al.}(2007)Eikerling, Kornyshev, and
  Kucernak]{Eikerling2007}
Eikerling,~M.~H.; Kornyshev,~A.~A.; Kucernak,~A. \emph{Physics World}
  \textbf{2007}, \emph{20}, 32--36\relax
\mciteBstWouldAddEndPuncttrue
\mciteSetBstMidEndSepPunct{\mcitedefaultmidpunct}
{\mcitedefaultendpunct}{\mcitedefaultseppunct}\relax
\EndOfBibitem
\bibitem[Vielstich \latin{et~al.}(2003)Vielstich, Lamm, and
  Gasteiger]{Vielstich2003}
Vielstich,~W.; Lamm,~A.; Gasteiger,~H.~A. In \emph{{Handbook of Fuel Cells:
  Fundamentals, Technology and Applications}}; Vielstich,~W., Lamm,~A.,
  Gasteiger,~H.~A., Eds.; WILEY, 2003\relax
\mciteBstWouldAddEndPuncttrue
\mciteSetBstMidEndSepPunct{\mcitedefaultmidpunct}
{\mcitedefaultendpunct}{\mcitedefaultseppunct}\relax
\EndOfBibitem
\bibitem[Weber and Newman(2004)Weber, and Newman]{Weber2004}
Weber,~A.~Z.; Newman,~J. \emph{Chemical Reviews} \textbf{2004}, \emph{104},
  4679--4726\relax
\mciteBstWouldAddEndPuncttrue
\mciteSetBstMidEndSepPunct{\mcitedefaultmidpunct}
{\mcitedefaultendpunct}{\mcitedefaultseppunct}\relax
\EndOfBibitem
\bibitem[Borup \latin{et~al.}(2007)Borup, Meyers, Pivovar, Kim, Mukundan,
  Garland, Myers, Wilson, Garzon, Wood, Zelenay, More, Stroh, Zawodzinski,
  Boncella, McGrath, Inaba, Miyatake, Hori, Ota, Ogumi, Miyata, Nishikata,
  Siroma, Uchimoto, Yasuda, Kimijima, and Iwashita]{Borup2007}
Borup,~R. \latin{et~al.}  \emph{Chemical Reviews} \textbf{2007}, \emph{107},
  3904--51\relax
\mciteBstWouldAddEndPuncttrue
\mciteSetBstMidEndSepPunct{\mcitedefaultmidpunct}
{\mcitedefaultendpunct}{\mcitedefaultseppunct}\relax
\EndOfBibitem
\bibitem[Peighambardoust \latin{et~al.}(2010)Peighambardoust, Rowshanzamira,
  and M.]{Peighambardoust2010}
Peighambardoust,~S.~J.; Rowshanzamira,~S.; M.,~A. \emph{International Journal
  of Hydrogen Energy} \textbf{2010}, \emph{35}, 9349--9384\relax
\mciteBstWouldAddEndPuncttrue
\mciteSetBstMidEndSepPunct{\mcitedefaultmidpunct}
{\mcitedefaultendpunct}{\mcitedefaultseppunct}\relax
\EndOfBibitem
\bibitem[Litster and McLean(2004)Litster, and McLean]{Litster2004}
Litster,~S.; McLean,~G. \emph{Journal of Power Sources} \textbf{2004},
  \emph{130}, 61--76\relax
\mciteBstWouldAddEndPuncttrue
\mciteSetBstMidEndSepPunct{\mcitedefaultmidpunct}
{\mcitedefaultendpunct}{\mcitedefaultseppunct}\relax
\EndOfBibitem
\bibitem[Mehta and Cooper(2003)Mehta, and Cooper]{Mehta2003}
Mehta,~V.; Cooper,~J.~S. \emph{Journal of Power Sources} \textbf{2003},
  \emph{114}, 32–53\relax
\mciteBstWouldAddEndPuncttrue
\mciteSetBstMidEndSepPunct{\mcitedefaultmidpunct}
{\mcitedefaultendpunct}{\mcitedefaultseppunct}\relax
\EndOfBibitem
\bibitem[Markovic and Jr.(2002)Markovic, and Jr.]{Markovic2002}
Markovic,~N.; Jr.,~P.~R. \emph{Surface Science Reports} \textbf{2002},
  \emph{45}, 117–229\relax
\mciteBstWouldAddEndPuncttrue
\mciteSetBstMidEndSepPunct{\mcitedefaultmidpunct}
{\mcitedefaultendpunct}{\mcitedefaultseppunct}\relax
\EndOfBibitem
\bibitem[Damjanovic and Brusic(1967)Damjanovic, and Brusic]{Damjanovic1967}
Damjanovic,~A.; Brusic,~V. \emph{Electrochimica Acta} \textbf{1967}, \emph{12},
  615--628\relax
\mciteBstWouldAddEndPuncttrue
\mciteSetBstMidEndSepPunct{\mcitedefaultmidpunct}
{\mcitedefaultendpunct}{\mcitedefaultseppunct}\relax
\EndOfBibitem
\bibitem[de~Morais \latin{et~al.}(2011)de~Morais, Sautet, Loffreda, and
  Franco]{DeMorais2011}
de~Morais,~R.~F.; Sautet,~P.; Loffreda,~D.; Franco,~A.~A. \emph{Electrochimica
  Acta} \textbf{2011}, \emph{56}, 10842--10856\relax
\mciteBstWouldAddEndPuncttrue
\mciteSetBstMidEndSepPunct{\mcitedefaultmidpunct}
{\mcitedefaultendpunct}{\mcitedefaultseppunct}\relax
\EndOfBibitem
\bibitem[Rinaldo \latin{et~al.}(2010)Rinaldo, Stumper, and
  Eikerling]{Rinaldo2010}
Rinaldo,~S.~G.; Stumper,~J.; Eikerling,~M. \emph{The journal of Physical
  Chemistry C} \textbf{2010}, \emph{114}, 5773--5785\relax
\mciteBstWouldAddEndPuncttrue
\mciteSetBstMidEndSepPunct{\mcitedefaultmidpunct}
{\mcitedefaultendpunct}{\mcitedefaultseppunct}\relax
\EndOfBibitem
\bibitem[Franco and Gerard(2008)Franco, and Gerard]{Franco2008}
Franco,~A.~A.; Gerard,~M. \emph{Journal of the Electrochemical Society}
  \textbf{2008}, \emph{155}, B367\relax
\mciteBstWouldAddEndPuncttrue
\mciteSetBstMidEndSepPunct{\mcitedefaultmidpunct}
{\mcitedefaultendpunct}{\mcitedefaultseppunct}\relax
\EndOfBibitem
\bibitem[Stamenkovic \latin{et~al.}(2007)Stamenkovic, Fowler, Mun, Wang, Ross,
  Lucas, and Marković]{Stamenkovic2007}
Stamenkovic,~V.~R.; Fowler,~B.; Mun,~B.~S.; Wang,~G.; Ross,~P.~N.;
  Lucas,~C.~A.; Marković,~N.~M. \emph{Science} \textbf{2007}, \emph{315},
  493--497\relax
\mciteBstWouldAddEndPuncttrue
\mciteSetBstMidEndSepPunct{\mcitedefaultmidpunct}
{\mcitedefaultendpunct}{\mcitedefaultseppunct}\relax
\EndOfBibitem
\bibitem[Gasteiger \latin{et~al.}(2005)Gasteiger, Kocha, Sompalli, and
  Wagner]{Gasteiger2005}
Gasteiger,~H.~A.; Kocha,~S.~S.; Sompalli,~B.; Wagner,~F.~T. \emph{Applied
  Catalysis B: Environmental} \textbf{2005}, \emph{56}, 9–35\relax
\mciteBstWouldAddEndPuncttrue
\mciteSetBstMidEndSepPunct{\mcitedefaultmidpunct}
{\mcitedefaultendpunct}{\mcitedefaultseppunct}\relax
\EndOfBibitem
\bibitem[Eikerling \latin{et~al.}(2007)Eikerling, Kornyshe, and
  Kulikovsky]{Eikerling2007b}
Eikerling,~M.; Kornyshe,~A.; Kulikovsky,~A. In \emph{{Physical modeling of fuel
  cells and their components, in Encyclopedia of electrochemistry}};
  Bard,~A.~J., Stratmann,~M., Macdonald,~D., Schmuki,~P., Eds.; Wiley-VCH,
  Weinheim, 2007\relax
\mciteBstWouldAddEndPuncttrue
\mciteSetBstMidEndSepPunct{\mcitedefaultmidpunct}
{\mcitedefaultendpunct}{\mcitedefaultseppunct}\relax
\EndOfBibitem
\bibitem[Eikerling and Malek(2009)Eikerling, and Malek]{Eikerling2009}
Eikerling,~M.~H.; Malek,~K. In \emph{{Proton Exchange Membrane Fuel Cells:
  Materials Properties and Performance}}; Wilkinson,~D.~P., Zhang,~J., Hui,~R.,
  Fergus,~J., Li,~X., Eds.; CRC Press, 2009; Chapter Physical Modeling of
  Materials for PEFCs: A Balancing Act of Water and Complex Morphologies, pp
  343--435\relax
\mciteBstWouldAddEndPuncttrue
\mciteSetBstMidEndSepPunct{\mcitedefaultmidpunct}
{\mcitedefaultendpunct}{\mcitedefaultseppunct}\relax
\EndOfBibitem
\bibitem[Malek \latin{et~al.}(2007)Malek, Eikerling, Wang, Navessin, and
  Liu]{Malek2007}
Malek,~K.; Eikerling,~M.; Wang,~Q.; Navessin,~T.; Liu,~Z. \emph{The Journal of
  Physical Chemistry C} \textbf{2007}, \emph{111}, 13627--13634\relax
\mciteBstWouldAddEndPuncttrue
\mciteSetBstMidEndSepPunct{\mcitedefaultmidpunct}
{\mcitedefaultendpunct}{\mcitedefaultseppunct}\relax
\EndOfBibitem
\bibitem[Malek and Franco(2011)Malek, and Franco]{Malek2011a}
Malek,~K.; Franco,~A.~A. \emph{The Journal of Physical Chemistry B}
  \textbf{2011}, \emph{115}, 8088--101\relax
\mciteBstWouldAddEndPuncttrue
\mciteSetBstMidEndSepPunct{\mcitedefaultmidpunct}
{\mcitedefaultendpunct}{\mcitedefaultseppunct}\relax
\EndOfBibitem
\bibitem[More \latin{et~al.}(2006)More, Borup, and Reeves]{More2006}
More,~K.; Borup,~R.; Reeves,~K. \emph{ECS Transactions} \textbf{2006},
  \emph{3}, 717--733\relax
\mciteBstWouldAddEndPuncttrue
\mciteSetBstMidEndSepPunct{\mcitedefaultmidpunct}
{\mcitedefaultendpunct}{\mcitedefaultseppunct}\relax
\EndOfBibitem
\bibitem[Xie \latin{et~al.}(2010)Xie, Xu, Wood, More, Zawodzinski, and
  Smith]{Xie2010}
Xie,~J.; Xu,~F.; Wood,~D.~L.; More,~K.~L.; Zawodzinski,~T.; Smith,~W.~H.
  \emph{Electrochimica Acta} \textbf{2010}, \emph{55}, 7404--7412\relax
\mciteBstWouldAddEndPuncttrue
\mciteSetBstMidEndSepPunct{\mcitedefaultmidpunct}
{\mcitedefaultendpunct}{\mcitedefaultseppunct}\relax
\EndOfBibitem
\bibitem[Wilson(1992)]{Wilson1992}
Wilson,~G.~S.,~M.~S. \emph{Journal of Applied Electrochemistry} \textbf{1992},
  \emph{22}, 1--7\relax
\mciteBstWouldAddEndPuncttrue
\mciteSetBstMidEndSepPunct{\mcitedefaultmidpunct}
{\mcitedefaultendpunct}{\mcitedefaultseppunct}\relax
\EndOfBibitem
\bibitem[Wilson(1993)]{Wilson1993}
Wilson,~M.~S. Membrane catalyst layer for fuel cells. 1993\relax
\mciteBstWouldAddEndPuncttrue
\mciteSetBstMidEndSepPunct{\mcitedefaultmidpunct}
{\mcitedefaultendpunct}{\mcitedefaultseppunct}\relax
\EndOfBibitem
\bibitem[Wilson \latin{et~al.}(1995)Wilson, Valerio, and
  Gottesfeld]{Wilson1995}
Wilson,~M.~S.; Valerio,~J.~A.; Gottesfeld,~S. \emph{Electrochimica Acta}
  \textbf{1995}, \emph{40}, 355--363\relax
\mciteBstWouldAddEndPuncttrue
\mciteSetBstMidEndSepPunct{\mcitedefaultmidpunct}
{\mcitedefaultendpunct}{\mcitedefaultseppunct}\relax
\EndOfBibitem
\bibitem[Li and Chan(2009)Li, and Chan]{Li2009}
Li,~A.; Chan,~S.~H. \emph{Electrochemistry Communications} \textbf{2009},
  \emph{11}, 897--900\relax
\mciteBstWouldAddEndPuncttrue
\mciteSetBstMidEndSepPunct{\mcitedefaultmidpunct}
{\mcitedefaultendpunct}{\mcitedefaultseppunct}\relax
\EndOfBibitem
\bibitem[Li \latin{et~al.}(2010)Li, Han, Chan, and Nguyen]{Li2010}
Li,~A.; Han,~M.; Chan,~S.~H.; Nguyen,~N.~T. \emph{Electrochimica Acta}
  \textbf{2010}, \emph{55}, 2706--2711\relax
\mciteBstWouldAddEndPuncttrue
\mciteSetBstMidEndSepPunct{\mcitedefaultmidpunct}
{\mcitedefaultendpunct}{\mcitedefaultseppunct}\relax
\EndOfBibitem
\bibitem[Mashio \latin{et~al.}(2010)Mashio, Malek, Eikerling, Ohma, Kanesaka,
  and Shinohara]{Mashio2010}
Mashio,~T.; Malek,~K.; Eikerling,~M.; Ohma,~A.; Kanesaka,~H.; Shinohara,~K.
  \emph{The Journal of Physical Chemistry C} \textbf{2010}, \emph{114},
  13739--13745\relax
\mciteBstWouldAddEndPuncttrue
\mciteSetBstMidEndSepPunct{\mcitedefaultmidpunct}
{\mcitedefaultendpunct}{\mcitedefaultseppunct}\relax
\EndOfBibitem
\bibitem[Chen \latin{et~al.}(2006)Chen, Sun, Guo, Zhao, Yan, Tian, Tang, Zhou,
  and Xin]{Chen2006}
Chen,~W.; Sun,~G.; Guo,~J.; Zhao,~X.; Yan,~S.; Tian,~J.; Tang,~S.; Zhou,~Z.;
  Xin,~Q. \emph{Electrochimica Acta} \textbf{2006}, \emph{51}, 2391--2399\relax
\mciteBstWouldAddEndPuncttrue
\mciteSetBstMidEndSepPunct{\mcitedefaultmidpunct}
{\mcitedefaultendpunct}{\mcitedefaultseppunct}\relax
\EndOfBibitem
\bibitem[Wang \latin{et~al.}(2009)Wang, Zuo, Chu, Shao, and Yin]{Wang2009}
Wang,~Z.~B.; Zuo,~P.~J.; Chu,~Y.~Y.; Shao,~Y.~Y.; Yin,~G.~P.
  \emph{International Journal of Hydrogen Energy} \textbf{2009}, \emph{34},
  4387--4394\relax
\mciteBstWouldAddEndPuncttrue
\mciteSetBstMidEndSepPunct{\mcitedefaultmidpunct}
{\mcitedefaultendpunct}{\mcitedefaultseppunct}\relax
\EndOfBibitem
\bibitem[Damasceno~Borges \latin{et~al.}(2013)Damasceno~Borges, Franco, Malek,
  Gebel, and Mossa]{Borges2013}
Damasceno~Borges,~D.; Franco,~A.~A.; Malek,~K.; Gebel,~G.; Mossa,~S. \emph{ACS
  nano} \textbf{2013}, \emph{7}, 6767--73\relax
\mciteBstWouldAddEndPuncttrue
\mciteSetBstMidEndSepPunct{\mcitedefaultmidpunct}
{\mcitedefaultendpunct}{\mcitedefaultseppunct}\relax
\EndOfBibitem
\bibitem[Damasceno~Borges \latin{et~al.}(2013)Damasceno~Borges, Malek, Mossa,
  Gebel, and Franco]{Borges2013b}
Damasceno~Borges,~D.; Malek,~K.; Mossa,~S.; Gebel,~G.; Franco,~A.~A. \emph{ECS
  - Transactions} \textbf{2013}, \emph{45}, 101--108\relax
\mciteBstWouldAddEndPuncttrue
\mciteSetBstMidEndSepPunct{\mcitedefaultmidpunct}
{\mcitedefaultendpunct}{\mcitedefaultseppunct}\relax
\EndOfBibitem
\bibitem[Mauritz and Moore(2004)Mauritz, and Moore]{Moore2004}
Mauritz,~K.~A.; Moore,~R.~B. \emph{Chemical Reviews} \textbf{2004}, \emph{104},
  4535--4585\relax
\mciteBstWouldAddEndPuncttrue
\mciteSetBstMidEndSepPunct{\mcitedefaultmidpunct}
{\mcitedefaultendpunct}{\mcitedefaultseppunct}\relax
\EndOfBibitem
\bibitem[Gierke \latin{et~al.}(1981)Gierke, Munn, and Wilson]{Gierke1981}
Gierke,~T.~D.; Munn,~G.; Wilson,~F.~C. \emph{Journal of Polymer Science:
  Polymer Physics Edition} \textbf{1981}, \emph{19}, 1687--1704\relax
\mciteBstWouldAddEndPuncttrue
\mciteSetBstMidEndSepPunct{\mcitedefaultmidpunct}
{\mcitedefaultendpunct}{\mcitedefaultseppunct}\relax
\EndOfBibitem
\bibitem[Hsu and Gierke(1983)Hsu, and Gierke]{Hsu1983}
Hsu,~W.~Y.; Gierke,~T.~D. \emph{Journal of Membrane Science} \textbf{1983},
  \emph{13}, 307--326\relax
\mciteBstWouldAddEndPuncttrue
\mciteSetBstMidEndSepPunct{\mcitedefaultmidpunct}
{\mcitedefaultendpunct}{\mcitedefaultseppunct}\relax
\EndOfBibitem
\bibitem[Yeager and Stek(1981)Yeager, and Stek]{Yeager1981}
Yeager,~H.~L.; Stek,~A. \emph{Journal of the Electrochemical Society}
  \textbf{1981}, \emph{128}, 1980--1984\relax
\mciteBstWouldAddEndPuncttrue
\mciteSetBstMidEndSepPunct{\mcitedefaultmidpunct}
{\mcitedefaultendpunct}{\mcitedefaultseppunct}\relax
\EndOfBibitem
\bibitem[Gebel \latin{et~al.}(1987)Gebel, Aldebert, and Pineri]{Gebel1987}
Gebel,~G.; Aldebert,~P.; Pineri,~M. \emph{Macromolecules} \textbf{1987},
  \emph{20}, 1425--1428\relax
\mciteBstWouldAddEndPuncttrue
\mciteSetBstMidEndSepPunct{\mcitedefaultmidpunct}
{\mcitedefaultendpunct}{\mcitedefaultseppunct}\relax
\EndOfBibitem
\bibitem[Gebel(2000)]{Gebel2000a}
Gebel,~G. \emph{Polymer} \textbf{2000}, \emph{41}, 5829--5838\relax
\mciteBstWouldAddEndPuncttrue
\mciteSetBstMidEndSepPunct{\mcitedefaultmidpunct}
{\mcitedefaultendpunct}{\mcitedefaultseppunct}\relax
\EndOfBibitem
\bibitem[Young \latin{et~al.}(2002)Young, Trevino, and Beck~Tan]{Young2002}
Young,~S.~K.; Trevino,~S.~F.; Beck~Tan,~N.~C. \emph{Journal of Polymer Science
  Part B: Polymer Physics} \textbf{2002}, \emph{40}, 387--400\relax
\mciteBstWouldAddEndPuncttrue
\mciteSetBstMidEndSepPunct{\mcitedefaultmidpunct}
{\mcitedefaultendpunct}{\mcitedefaultseppunct}\relax
\EndOfBibitem
\bibitem[Rubatat \latin{et~al.}(2002)Rubatat, Rollet, Gebel, and
  Diat]{Rubatat2002}
Rubatat,~L.; Rollet,~A.~L.; Gebel,~G.; Diat,~O. \emph{Macromolecules}
  \textbf{2002}, \emph{35}, 4050--4055\relax
\mciteBstWouldAddEndPuncttrue
\mciteSetBstMidEndSepPunct{\mcitedefaultmidpunct}
{\mcitedefaultendpunct}{\mcitedefaultseppunct}\relax
\EndOfBibitem
\bibitem[Schmidt-Rohr and Chen(2008)Schmidt-Rohr, and Chen]{Schmidt-Rohr2008}
Schmidt-Rohr,~K.; Chen,~Q. \emph{Nature Materials} \textbf{2008}, \emph{7},
  75--83\relax
\mciteBstWouldAddEndPuncttrue
\mciteSetBstMidEndSepPunct{\mcitedefaultmidpunct}
{\mcitedefaultendpunct}{\mcitedefaultseppunct}\relax
\EndOfBibitem
\bibitem[Elliott \latin{et~al.}(2011)Elliott, Wu, Paddison, and
  Moore]{Elliott2011}
Elliott,~J.~A.; Wu,~D.; Paddison,~S.~J.; Moore,~R.~B. \emph{Soft Matter}
  \textbf{2011}, \emph{7}, 6820--6827\relax
\mciteBstWouldAddEndPuncttrue
\mciteSetBstMidEndSepPunct{\mcitedefaultmidpunct}
{\mcitedefaultendpunct}{\mcitedefaultseppunct}\relax
\EndOfBibitem
\bibitem[Ma \latin{et~al.}(2007)Ma, Chen, Jogensen, Stein, and Skou]{Ma2007}
Ma,~S.; Chen,~Q.; Jogensen,~F.; Stein,~P.; Skou,~E. \emph{Solid State Ionics}
  \textbf{2007}, \emph{178}, 1568--1575\relax
\mciteBstWouldAddEndPuncttrue
\mciteSetBstMidEndSepPunct{\mcitedefaultmidpunct}
{\mcitedefaultendpunct}{\mcitedefaultseppunct}\relax
\EndOfBibitem
\bibitem[Paul \latin{et~al.}(2011)Paul, Fraser, Pearce, and Karan]{Paul2011}
Paul,~D.~K.; Fraser,~A.; Pearce,~J.; Karan,~K. \emph{ECS Transactions}
  \textbf{2011}, \emph{41}, 1393--1406\relax
\mciteBstWouldAddEndPuncttrue
\mciteSetBstMidEndSepPunct{\mcitedefaultmidpunct}
{\mcitedefaultendpunct}{\mcitedefaultseppunct}\relax
\EndOfBibitem
\bibitem[Paul \latin{et~al.}(2011)Paul, Fraser, and Karan]{Paul2011a}
Paul,~D.~K.; Fraser,~A.; Karan,~K. \emph{Electrochemistry Communications}
  \textbf{2011}, \emph{13}, 774--777\relax
\mciteBstWouldAddEndPuncttrue
\mciteSetBstMidEndSepPunct{\mcitedefaultmidpunct}
{\mcitedefaultendpunct}{\mcitedefaultseppunct}\relax
\EndOfBibitem
\bibitem[Paul \latin{et~al.}(2013)Paul, Karan, Docoslis, Giorgi, and
  Pearce]{Paul2013}
Paul,~D.~K.; Karan,~K.; Docoslis,~A.; Giorgi,~J.~B.; Pearce,~J.
  \emph{Macromolecules} \textbf{2013}, \emph{46}, 3461--3475\relax
\mciteBstWouldAddEndPuncttrue
\mciteSetBstMidEndSepPunct{\mcitedefaultmidpunct}
{\mcitedefaultendpunct}{\mcitedefaultseppunct}\relax
\EndOfBibitem
\bibitem[Wood \latin{et~al.}(2009)Wood, Chlistunoff, Majewski, and
  Borup]{Wood2009}
Wood,~D.~L.; Chlistunoff,~J.; Majewski,~J.; Borup,~R.~L. \emph{Journal of the
  American Chemical Society} \textbf{2009}, \emph{131}, 18096--104\relax
\mciteBstWouldAddEndPuncttrue
\mciteSetBstMidEndSepPunct{\mcitedefaultmidpunct}
{\mcitedefaultendpunct}{\mcitedefaultseppunct}\relax
\EndOfBibitem
\bibitem[Dura \latin{et~al.}(2009)Dura, Murthi, Hartman, Satija, and
  Majkrzak]{Dura2009}
Dura,~J.~A.; Murthi,~V.~S.; Hartman,~M.; Satija,~S.~K.; Majkrzak,~C.~F.
  \emph{Macromolecules} \textbf{2009}, \emph{42}, 4769--4774\relax
\mciteBstWouldAddEndPuncttrue
\mciteSetBstMidEndSepPunct{\mcitedefaultmidpunct}
{\mcitedefaultendpunct}{\mcitedefaultseppunct}\relax
\EndOfBibitem
\bibitem[Masuda \latin{et~al.}(2009)Masuda, Naohara, Takakusagi, Singh, and
  Uosaki]{Masuda2009}
Masuda,~T.; Naohara,~H.; Takakusagi,~S.; Singh,~P.~R.; Uosaki,~K.
  \emph{Chemistry Letters} \textbf{2009}, \emph{38}, 884--885\relax
\mciteBstWouldAddEndPuncttrue
\mciteSetBstMidEndSepPunct{\mcitedefaultmidpunct}
{\mcitedefaultendpunct}{\mcitedefaultseppunct}\relax
\EndOfBibitem
\bibitem[Koestner \latin{et~al.}(2011)Koestner, Roiter, Kozhinova, and
  Minko]{Koestner2011}
Koestner,~R.; Roiter,~Y.; Kozhinova,~I.; Minko,~S. \emph{Langmuir}
  \textbf{2011}, \emph{27}, 10157--10166\relax
\mciteBstWouldAddEndPuncttrue
\mciteSetBstMidEndSepPunct{\mcitedefaultmidpunct}
{\mcitedefaultendpunct}{\mcitedefaultseppunct}\relax
\EndOfBibitem
\bibitem[Eastman \latin{et~al.}(2012)Eastman, Kim, Page, Kang, and
  Soles]{Eastman2012}
Eastman,~S.~A.; Kim,~S.; Page,~B.~W.,~K. A.and~Rowe; Kang,~S.; Soles,~C.~L.
  \emph{Macromolecules} \textbf{2012}, \emph{45}, 7920--7930\relax
\mciteBstWouldAddEndPuncttrue
\mciteSetBstMidEndSepPunct{\mcitedefaultmidpunct}
{\mcitedefaultendpunct}{\mcitedefaultseppunct}\relax
\EndOfBibitem
\bibitem[Nagao(2013)]{Nagao2013}
Nagao,~Y. \emph{The Journal of Physical Chemistry C} \textbf{2013}, \emph{117},
  3294--3297\relax
\mciteBstWouldAddEndPuncttrue
\mciteSetBstMidEndSepPunct{\mcitedefaultmidpunct}
{\mcitedefaultendpunct}{\mcitedefaultseppunct}\relax
\EndOfBibitem
\bibitem[Kusoglu \latin{et~al.}(2014)Kusoglu, Kushner, Paul, Karan, Hickner,
  and Weber]{Kusoglu2014}
Kusoglu,~A.; Kushner,~D.; Paul,~D.~K.; Karan,~K.; Hickner,~M.~A.; Weber,~A.~Z.
  \emph{Advanced Functional Materials} \textbf{2014}, \relax
\mciteBstWouldAddEndPunctfalse
\mciteSetBstMidEndSepPunct{\mcitedefaultmidpunct}
{}{\mcitedefaultseppunct}\relax
\EndOfBibitem
\bibitem[Modestino \latin{et~al.}(2012)Modestino, Kusoglu, Hexemer, Weber, and
  Segalman]{Modestino2012}
Modestino,~M.~A.; Kusoglu,~A.; Hexemer,~A.; Weber,~A.~Z.; Segalman,~R.~A.
  \emph{Macromolecules} \textbf{2012}, \emph{45}, 4681--4688\relax
\mciteBstWouldAddEndPuncttrue
\mciteSetBstMidEndSepPunct{\mcitedefaultmidpunct}
{\mcitedefaultendpunct}{\mcitedefaultseppunct}\relax
\EndOfBibitem
\bibitem[Modestino \latin{et~al.}(2013)Modestino, Paul, Dishari, Petrina,
  Allen, Hickner, Karan, Segalman, and Weber]{Modestino2013}
Modestino,~M.~A.; Paul,~D.~K.; Dishari,~S.; Petrina,~S.; Allen,~F.;
  Hickner,~M.; Karan,~K.; Segalman,~R.~A.; Weber,~A.~Z. \emph{Macromolecules}
  \textbf{2013}, \emph{46}, 867−873\relax
\mciteBstWouldAddEndPuncttrue
\mciteSetBstMidEndSepPunct{\mcitedefaultmidpunct}
{\mcitedefaultendpunct}{\mcitedefaultseppunct}\relax
\EndOfBibitem
\bibitem[Bass \latin{et~al.}(2010)Bass, Berman, Singh, Konovalov, and
  Freger]{Bass2010}
Bass,~M.; Berman,~A.; Singh,~A.; Konovalov,~O.; Freger,~V. \emph{The Journal of
  Physical Chemistry B} \textbf{2010}, \emph{114}, 3784--90\relax
\mciteBstWouldAddEndPuncttrue
\mciteSetBstMidEndSepPunct{\mcitedefaultmidpunct}
{\mcitedefaultendpunct}{\mcitedefaultseppunct}\relax
\EndOfBibitem
\bibitem[Bass \latin{et~al.}(2011)Bass, Berman, Singh, Konovalov, and
  Freger]{Bass2011}
Bass,~M.; Berman,~A.; Singh,~A.; Konovalov,~O.; Freger,~V.
  \emph{Macromolecules} \textbf{2011}, \emph{44}, 2893--2899\relax
\mciteBstWouldAddEndPuncttrue
\mciteSetBstMidEndSepPunct{\mcitedefaultmidpunct}
{\mcitedefaultendpunct}{\mcitedefaultseppunct}\relax
\EndOfBibitem
\bibitem[Balbuena \latin{et~al.}(2005)Balbuena, Lamas, and Wang]{Balbuena2005}
Balbuena,~P.; Lamas,~E.; Wang,~Y. \emph{Electrochimica Acta} \textbf{2005},
  \emph{50}, 3788--3795\relax
\mciteBstWouldAddEndPuncttrue
\mciteSetBstMidEndSepPunct{\mcitedefaultmidpunct}
{\mcitedefaultendpunct}{\mcitedefaultseppunct}\relax
\EndOfBibitem
\bibitem[Lamas and Balbuena(2006)Lamas, and Balbuena]{Lamas2006}
Lamas,~E.; Balbuena,~P. \emph{Electrochimica Acta} \textbf{2006}, \emph{51},
  5904--5911\relax
\mciteBstWouldAddEndPuncttrue
\mciteSetBstMidEndSepPunct{\mcitedefaultmidpunct}
{\mcitedefaultendpunct}{\mcitedefaultseppunct}\relax
\EndOfBibitem
\bibitem[Liu \latin{et~al.}(2008)Liu, Selvan, Cui, Edwards, Keffer, and
  Steele]{Liu2008}
Liu,~J.; Selvan,~M.~E.; Cui,~S.; Edwards,~B.~J.; Keffer,~D.~J.; Steele,~W.~V.
  \emph{The Journal of Physical Chemistry C} \textbf{2008}, \emph{112},
  1985--1993\relax
\mciteBstWouldAddEndPuncttrue
\mciteSetBstMidEndSepPunct{\mcitedefaultmidpunct}
{\mcitedefaultendpunct}{\mcitedefaultseppunct}\relax
\EndOfBibitem
\bibitem[Selvan \latin{et~al.}(2012)Selvan, He, Calvo-mun, and
  Keffer]{Selvan2012}
Selvan,~M.~E.; He,~Q.; Calvo-mun,~E.~M.; Keffer,~D.~J. \emph{The Journal of
  Physical Chemistry C} \textbf{2012}, \emph{116}, 12890--12899\relax
\mciteBstWouldAddEndPuncttrue
\mciteSetBstMidEndSepPunct{\mcitedefaultmidpunct}
{\mcitedefaultendpunct}{\mcitedefaultseppunct}\relax
\EndOfBibitem
\bibitem[Selvan \latin{et~al.}(2008)Selvan, Liu, Keffer, Cui, Edwards, and
  Steele]{Selvan2008}
Selvan,~M.~E.; Liu,~J.; Keffer,~D.~J.; Cui,~S.; Edwards,~B.~J.; Steele,~W.~V.
  \emph{The Journal of Physical Chemistry C} \textbf{2008}, \emph{112},
  1975--1984\relax
\mciteBstWouldAddEndPuncttrue
\mciteSetBstMidEndSepPunct{\mcitedefaultmidpunct}
{\mcitedefaultendpunct}{\mcitedefaultseppunct}\relax
\EndOfBibitem
\bibitem[Urata \latin{et~al.}(2005)Urata, Irisawa, Takada, Shinoda, Tsuzuki,
  and Mikami]{Urata2005}
Urata,~S.; Irisawa,~J.; Takada,~A.; Shinoda,~W.; Tsuzuki,~S.; Mikami,~M.
  \emph{The Journal of Physical Chemistry B} \textbf{2005}, \emph{109},
  4269--78\relax
\mciteBstWouldAddEndPuncttrue
\mciteSetBstMidEndSepPunct{\mcitedefaultmidpunct}
{\mcitedefaultendpunct}{\mcitedefaultseppunct}\relax
\EndOfBibitem
\bibitem[Allahyarov and Taylor(2007)Allahyarov, and Taylor]{Allahyarov2007}
Allahyarov,~E.; Taylor,~P.~L. \emph{Journal of Chemical Physics} \textbf{2007},
  \emph{127}, 154901\relax
\mciteBstWouldAddEndPuncttrue
\mciteSetBstMidEndSepPunct{\mcitedefaultmidpunct}
{\mcitedefaultendpunct}{\mcitedefaultseppunct}\relax
\EndOfBibitem
\bibitem[Allahyarov \latin{et~al.}(2009)Allahyarov, Taylor, and
  L\"{o}wen]{Allahyarov2009}
Allahyarov,~E.; Taylor,~P.~L.; L\"{o}wen,~H. \emph{Physical Review E}
  \textbf{2009}, \emph{80}, 061802\relax
\mciteBstWouldAddEndPuncttrue
\mciteSetBstMidEndSepPunct{\mcitedefaultmidpunct}
{\mcitedefaultendpunct}{\mcitedefaultseppunct}\relax
\EndOfBibitem
\bibitem[Cui \latin{et~al.}(2007)Cui, Liu, Selvan, Keffer, Edwards, and
  Steele]{Cui2007}
Cui,~S.; Liu,~J.; Selvan,~M.~E.; Keffer,~D.~J.; Edwards,~B.~J.; Steele,~W.~V.
  \emph{The Journal of Physical Chemistry B} \textbf{2007}, \emph{111},
  2208--2218\relax
\mciteBstWouldAddEndPuncttrue
\mciteSetBstMidEndSepPunct{\mcitedefaultmidpunct}
{\mcitedefaultendpunct}{\mcitedefaultseppunct}\relax
\EndOfBibitem
\bibitem[Cui \latin{et~al.}(2008)Cui, Liu, Selvan, Paddison, Keffer, Edwards,
  and Steele]{Cui2008}
Cui,~S.; Liu,~J.; Selvan,~M.~E.; Paddison,~S.~J.; Keffer,~D.~J.;
  Edwards,~B.~J.; Steele,~W.~V. \emph{The Journal of Physical Chemistry B}
  \textbf{2008}, \emph{112}, 13273--13284\relax
\mciteBstWouldAddEndPuncttrue
\mciteSetBstMidEndSepPunct{\mcitedefaultmidpunct}
{\mcitedefaultendpunct}{\mcitedefaultseppunct}\relax
\EndOfBibitem
\bibitem[Vishnyakov and Neimark(2000)Vishnyakov, and Neimark]{Vishnyakov2000}
Vishnyakov,~A.; Neimark,~A.~V. \emph{The Journal of Physical Chemistry B}
  \textbf{2000}, \emph{104}, 4471--4478\relax
\mciteBstWouldAddEndPuncttrue
\mciteSetBstMidEndSepPunct{\mcitedefaultmidpunct}
{\mcitedefaultendpunct}{\mcitedefaultseppunct}\relax
\EndOfBibitem
\bibitem[Vishnyakov and Neimark(2001)Vishnyakov, and Neimark]{Vishnyakov2001}
Vishnyakov,~A.; Neimark,~A.~V. \emph{The Journal of Physical Chemistry B}
  \textbf{2001}, \emph{105}, 9586--9594\relax
\mciteBstWouldAddEndPuncttrue
\mciteSetBstMidEndSepPunct{\mcitedefaultmidpunct}
{\mcitedefaultendpunct}{\mcitedefaultseppunct}\relax
\EndOfBibitem
\bibitem[Liu \latin{et~al.}(2010)Liu, Suraweera, Keffer, Cui, and
  Paddison]{Liu2010}
Liu,~J.; Suraweera,~S.; Keffer,~D.~J.; Cui,~S.; Paddison,~S.~J. \emph{Journal
  Physical Chemistry C} \textbf{2010}, \emph{114}, 11279--11292\relax
\mciteBstWouldAddEndPuncttrue
\mciteSetBstMidEndSepPunct{\mcitedefaultmidpunct}
{\mcitedefaultendpunct}{\mcitedefaultseppunct}\relax
\EndOfBibitem
\bibitem[Venkatnathan \latin{et~al.}(2007)Venkatnathan, Devanathan, and
  Dupuis]{Venkatnathan2007}
Venkatnathan,~A.; Devanathan,~R.; Dupuis,~M. \emph{The Journal of Physical
  Chemistry B} \textbf{2007}, \emph{111}, 7234--7244\relax
\mciteBstWouldAddEndPuncttrue
\mciteSetBstMidEndSepPunct{\mcitedefaultmidpunct}
{\mcitedefaultendpunct}{\mcitedefaultseppunct}\relax
\EndOfBibitem
\bibitem[Kusaka \latin{et~al.}(1998)Kusaka, G., and H.]{Kusaka1998}
Kusaka,~I.; G.,~W.~Z.; H.,~S.~J. \emph{Journal of Chemical Physics}
  \textbf{1998}, \emph{108}, 6829\relax
\mciteBstWouldAddEndPuncttrue
\mciteSetBstMidEndSepPunct{\mcitedefaultmidpunct}
{\mcitedefaultendpunct}{\mcitedefaultseppunct}\relax
\EndOfBibitem
\bibitem[Berendsen \latin{et~al.}(1987)Berendsen, Grigera, and
  Straatsma]{Berendsen1987}
Berendsen,~H. J.~C.; Grigera,~J.~R.; Straatsma,~T.~P. \emph{The Journal of
  Physical Chemistry} \textbf{1987}, \emph{91}, 6269--6271\relax
\mciteBstWouldAddEndPuncttrue
\mciteSetBstMidEndSepPunct{\mcitedefaultmidpunct}
{\mcitedefaultendpunct}{\mcitedefaultseppunct}\relax
\EndOfBibitem
\bibitem[Giovambattista \latin{et~al.}(2007)Giovambattista, Debenedetti, and
  Rossky]{Giovambattista2007}
Giovambattista,~N.; Debenedetti,~P.~G.; Rossky,~P.~J. \emph{The Journal of
  Physical Chemistry B} \textbf{2007}, \emph{111}, 9581--9587\relax
\mciteBstWouldAddEndPuncttrue
\mciteSetBstMidEndSepPunct{\mcitedefaultmidpunct}
{\mcitedefaultendpunct}{\mcitedefaultseppunct}\relax
\EndOfBibitem
\bibitem[Castrill\'on \latin{et~al.}(2009)Castrill\'on, Giovambattista, Aksay,
  and Debenedetti]{Castrillon2009}
Castrill\'on,~S. R.-V.; Giovambattista,~N.; Aksay,~I.; Debenedetti,~P.~G.
  \emph{The Journal of Physical Chemistry B} \textbf{2009}, \emph{113},
  1438--1446\relax
\mciteBstWouldAddEndPuncttrue
\mciteSetBstMidEndSepPunct{\mcitedefaultmidpunct}
{\mcitedefaultendpunct}{\mcitedefaultseppunct}\relax
\EndOfBibitem
\bibitem[Nijmeijer \latin{et~al.}(1990)Nijmeijer, Bruin, Bakker, and van
  Leeuwen]{Nijmeijer1990}
Nijmeijer,~M. J.~P.; Bruin,~C.; Bakker,~A.~F.; van Leeuwen,~J. M.~J.
  \emph{Physical Review} \textbf{1990}, \emph{42}, 6052\relax
\mciteBstWouldAddEndPuncttrue
\mciteSetBstMidEndSepPunct{\mcitedefaultmidpunct}
{\mcitedefaultendpunct}{\mcitedefaultseppunct}\relax
\EndOfBibitem
\bibitem[Scheidler \latin{et~al.}(2002)Scheidler, Kob, and
  Binder]{Scheidler2002}
Scheidler,~P.; Kob,~W.; Binder,~K. \emph{Europhysics Letters} \textbf{2002},
  \emph{59}, 701--707\relax
\mciteBstWouldAddEndPuncttrue
\mciteSetBstMidEndSepPunct{\mcitedefaultmidpunct}
{\mcitedefaultendpunct}{\mcitedefaultseppunct}\relax
\EndOfBibitem
\bibitem[Spohr and Heinzinger(1988)Spohr, and Heinzinger]{Spohr1988}
Spohr,~E.; Heinzinger,~K. \emph{Electrochimica Acta} \textbf{1988}, \emph{33},
  1211--222\relax
\mciteBstWouldAddEndPuncttrue
\mciteSetBstMidEndSepPunct{\mcitedefaultmidpunct}
{\mcitedefaultendpunct}{\mcitedefaultseppunct}\relax
\EndOfBibitem
\bibitem[Abraham and Singh(1977)Abraham, and Singh]{Abraham1977}
Abraham,~F.~F.; Singh,~Y. \emph{Journal of Chemical Physics} \textbf{1977},
  \emph{67}, 2384\relax
\mciteBstWouldAddEndPuncttrue
\mciteSetBstMidEndSepPunct{\mcitedefaultmidpunct}
{\mcitedefaultendpunct}{\mcitedefaultseppunct}\relax
\EndOfBibitem
\bibitem[Shi and Dhir(2009)Shi, and Dhir]{Shi2009}
Shi,~B.; Dhir,~V.~K. \emph{Journal of Chemical Physics} \textbf{2009},
  \emph{130}, 034705\relax
\mciteBstWouldAddEndPuncttrue
\mciteSetBstMidEndSepPunct{\mcitedefaultmidpunct}
{\mcitedefaultendpunct}{\mcitedefaultseppunct}\relax
\EndOfBibitem
\bibitem[Werder \latin{et~al.}(2003)Werder, Walther, Jaffe, Halicioglu, and
  Koumoutsakos]{Werder2003}
Werder,~T.; Walther,~J.~H.; Jaffe,~R.~L.; Halicioglu,~T.; Koumoutsakos,~P.
  \emph{The Journal of Physical Chemistry B} \textbf{2003}, \emph{107},
  1345--1352\relax
\mciteBstWouldAddEndPuncttrue
\mciteSetBstMidEndSepPunct{\mcitedefaultmidpunct}
{\mcitedefaultendpunct}{\mcitedefaultseppunct}\relax
\EndOfBibitem
\bibitem[Werder \latin{et~al.}(2001)Werder, Walther, Jaffe, Halicioglu, Noca,
  and Koumoutsakos]{Werder2001}
Werder,~T.; Walther,~J.~H.; Jaffe,~R.~L.; Halicioglu,~T.; Noca,~F.;
  Koumoutsakos,~P. \emph{Nano Letters} \textbf{2001}, \emph{1}, 697--702\relax
\mciteBstWouldAddEndPuncttrue
\mciteSetBstMidEndSepPunct{\mcitedefaultmidpunct}
{\mcitedefaultendpunct}{\mcitedefaultseppunct}\relax
\EndOfBibitem
\bibitem[Spohr(1989)]{Spohr1989}
Spohr,~E. \emph{The Journal of Physical Chemistry} \textbf{1989}, \emph{93},
  6171--6180\relax
\mciteBstWouldAddEndPuncttrue
\mciteSetBstMidEndSepPunct{\mcitedefaultmidpunct}
{\mcitedefaultendpunct}{\mcitedefaultseppunct}\relax
\EndOfBibitem
\bibitem[Lee and Rossky(1994)Lee, and Rossky]{Lee1994}
Lee,~S.~H.; Rossky,~P.~J. \emph{Journal of Chemical Physics} \textbf{1994},
  \emph{100}, 3334\relax
\mciteBstWouldAddEndPuncttrue
\mciteSetBstMidEndSepPunct{\mcitedefaultmidpunct}
{\mcitedefaultendpunct}{\mcitedefaultseppunct}\relax
\EndOfBibitem
\bibitem[Glebov \latin{et~al.}(1997)Glebov, Graham, Menzel, and
  Toennies]{Glebov1997}
Glebov,~A.; Graham,~A.~P.; Menzel,~A.; Toennies,~J.~P. \emph{Journal of
  Chemical Physics} \textbf{1997}, \emph{106}, 9382\relax
\mciteBstWouldAddEndPuncttrue
\mciteSetBstMidEndSepPunct{\mcitedefaultmidpunct}
{\mcitedefaultendpunct}{\mcitedefaultseppunct}\relax
\EndOfBibitem
\bibitem[Tatarkhanov \latin{et~al.}(2009)Tatarkhanov, Ogletree, Rose, Mitsui,
  Fomin, Maier, Rose, Cerd\'a, and Salmeron]{Tatarkhanov2009}
Tatarkhanov,~M.; Ogletree,~D.~F.; Rose,~F.; Mitsui,~T.; Fomin,~E.; Maier,~S.;
  Rose,~M.; Cerd\'a,~J.~I.; Salmeron,~M. \emph{Journal of the American Chemical
  Society} \textbf{2009}, \emph{131}, 18425--18434\relax
\mciteBstWouldAddEndPuncttrue
\mciteSetBstMidEndSepPunct{\mcitedefaultmidpunct}
{\mcitedefaultendpunct}{\mcitedefaultseppunct}\relax
\EndOfBibitem
\bibitem[Cheng \latin{et~al.}(2010)Cheng, Malek, Sui, and Djilali]{Cheng2010}
Cheng,~C.; Malek,~K.; Sui,~P.; Djilali,~N. \emph{Electrochimica Acta}
  \textbf{2010}, \emph{55}, 1588--1597\relax
\mciteBstWouldAddEndPuncttrue
\mciteSetBstMidEndSepPunct{\mcitedefaultmidpunct}
{\mcitedefaultendpunct}{\mcitedefaultseppunct}\relax
\EndOfBibitem
\bibitem[Kreuer(2000)]{Kreuer2000}
Kreuer,~K.~D. \emph{Solid State Ionics} \textbf{2000}, \emph{136-137},
  149--160\relax
\mciteBstWouldAddEndPuncttrue
\mciteSetBstMidEndSepPunct{\mcitedefaultmidpunct}
{\mcitedefaultendpunct}{\mcitedefaultseppunct}\relax
\EndOfBibitem
\bibitem[Elliott and Paddison(2007)Elliott, and Paddison]{Elliott2007}
Elliott,~J.~A.; Paddison,~S.~J. \emph{Physical Chemistry Chemical Physics}
  \textbf{2007}, \emph{9}, 2602--18\relax
\mciteBstWouldAddEndPuncttrue
\mciteSetBstMidEndSepPunct{\mcitedefaultmidpunct}
{\mcitedefaultendpunct}{\mcitedefaultseppunct}\relax
\EndOfBibitem
\bibitem[Subbaraman \latin{et~al.}(2010)Subbaraman, Strmcnik, Stamenkovic, and
  Markovic]{Subbaraman2010}
Subbaraman,~R.; Strmcnik,~D.; Stamenkovic,~V.; Markovic,~N.~M. \emph{The
  Journal of Physical Chemistry C} \textbf{2010}, \emph{114}, 8414--8422\relax
\mciteBstWouldAddEndPuncttrue
\mciteSetBstMidEndSepPunct{\mcitedefaultmidpunct}
{\mcitedefaultendpunct}{\mcitedefaultseppunct}\relax
\EndOfBibitem
\bibitem[Franco \latin{et~al.}(2006)Franco, Schott, Jallut, and
  Maschke]{Franco2006}
Franco,~A.~A.; Schott,~P.; Jallut,~C.; Maschke,~B. \emph{Journal of the
  Electrochemical Society} \textbf{2006}, \emph{153}, A1053--A1061\relax
\mciteBstWouldAddEndPuncttrue
\mciteSetBstMidEndSepPunct{\mcitedefaultmidpunct}
{\mcitedefaultendpunct}{\mcitedefaultseppunct}\relax
\EndOfBibitem
\bibitem[Subbaraman \latin{et~al.}(2010)Subbaraman, Strmcnik, Paulikas,
  Stamenkovic, and Markovic]{Subbaraman2010a}
Subbaraman,~R.; Strmcnik,~D.; Paulikas,~A.~P.; Stamenkovic,~V.~R.;
  Markovic,~N.~M. \emph{ChemPhysChem} \textbf{2010}, \emph{11}, 2825--33\relax
\mciteBstWouldAddEndPuncttrue
\mciteSetBstMidEndSepPunct{\mcitedefaultmidpunct}
{\mcitedefaultendpunct}{\mcitedefaultseppunct}\relax
\EndOfBibitem
\bibitem[Quiroga \latin{et~al.}(2014)Quiroga, Xue, Nguyen, Huang, Tulodziecki,
  and Franco]{Quiroga2014}
Quiroga,~M.; Xue,~K.; Nguyen,~T.; Huang,~H.; Tulodziecki,~M.; Franco,~A.
  \emph{Journal of the Electrochemical Society} \textbf{2014}, \emph{161},
  E3302--E3310\relax
\mciteBstWouldAddEndPuncttrue
\mciteSetBstMidEndSepPunct{\mcitedefaultmidpunct}
{\mcitedefaultendpunct}{\mcitedefaultseppunct}\relax
\EndOfBibitem
\bibitem[Wang \latin{et~al.}(2009)Wang, Roudgar, and
  Eikerling]{WangLRoudgarA2009}
Wang,~L.; Roudgar,~A.; Eikerling,~M. \emph{The Journal of Physical Chemistry C}
  \textbf{2009}, \emph{113}, 17989--17996\relax
\mciteBstWouldAddEndPuncttrue
\mciteSetBstMidEndSepPunct{\mcitedefaultmidpunct}
{\mcitedefaultendpunct}{\mcitedefaultseppunct}\relax
\EndOfBibitem
\bibitem[Krapf \latin{et~al.}(2006)Krapf, Quinn, Wu, Zandbergen, Dekker, and
  Lemay]{Krapf2006}
Krapf,~D.; Quinn,~B.~M.; Wu,~M.~Y.; Zandbergen,~H.~W.; Dekker,~C.; Lemay,~S.~G.
  \emph{Nano Letters} \textbf{2006}, \emph{6}, 2531--2535\relax
\mciteBstWouldAddEndPuncttrue
\mciteSetBstMidEndSepPunct{\mcitedefaultmidpunct}
{\mcitedefaultendpunct}{\mcitedefaultseppunct}\relax
\EndOfBibitem
\bibitem[Zhdanov and Kasemo(2006)Zhdanov, and Kasemo]{Zhdanov2006561}
Zhdanov,~V.~P.; Kasemo,~B. \emph{Electrochemistry Communications}
  \textbf{2006}, \emph{8}, 561--564\relax
\mciteBstWouldAddEndPuncttrue
\mciteSetBstMidEndSepPunct{\mcitedefaultmidpunct}
{\mcitedefaultendpunct}{\mcitedefaultseppunct}\relax
\EndOfBibitem
\bibitem[Zhdanov and Kasemo(2004)Zhdanov, and Kasemo]{Zhdanov2004}
Zhdanov,~V.~p.; Kasemo,~B. \emph{Surface Science} \textbf{2004}, \emph{554},
  103--108\relax
\mciteBstWouldAddEndPuncttrue
\mciteSetBstMidEndSepPunct{\mcitedefaultmidpunct}
{\mcitedefaultendpunct}{\mcitedefaultseppunct}\relax
\EndOfBibitem
\bibitem[Zhdanov and Kasemo(2008)Zhdanov, and Kasemo]{Zhdanov2008}
Zhdanov,~V.~P.; Kasemo,~B. \emph{Chemical Physics Letters} \textbf{2008},
  \emph{460}, 158--161\relax
\mciteBstWouldAddEndPuncttrue
\mciteSetBstMidEndSepPunct{\mcitedefaultmidpunct}
{\mcitedefaultendpunct}{\mcitedefaultseppunct}\relax
\EndOfBibitem
\bibitem[Biesheuvel \latin{et~al.}(2009)Biesheuvel, Franco, and
  Bazant]{Biesheuvel2009}
Biesheuvel,~P.~M.; Franco,~A.~A.; Bazant,~M.~Z. \emph{Journal of The
  Electrochemical Society} \textbf{2009}, \emph{156}, B225\relax
\mciteBstWouldAddEndPuncttrue
\mciteSetBstMidEndSepPunct{\mcitedefaultmidpunct}
{\mcitedefaultendpunct}{\mcitedefaultseppunct}\relax
\EndOfBibitem
\bibitem[Freger(2009)]{Freger2009}
Freger,~V. \emph{The Journal of Physical Chemistry B} \textbf{2009},
  \emph{113}, 24--36\relax
\mciteBstWouldAddEndPuncttrue
\mciteSetBstMidEndSepPunct{\mcitedefaultmidpunct}
{\mcitedefaultendpunct}{\mcitedefaultseppunct}\relax
\EndOfBibitem
\bibitem[Choi and Datta(2003)Choi, and Datta]{Choi2003}
Choi,~P.; Datta,~R. \emph{Journal of The Electrochemical Society}
  \textbf{2003}, \emph{150}, E601\relax
\mciteBstWouldAddEndPuncttrue
\mciteSetBstMidEndSepPunct{\mcitedefaultmidpunct}
{\mcitedefaultendpunct}{\mcitedefaultseppunct}\relax
\EndOfBibitem
\bibitem[Huang \latin{et~al.}(2012)Huang, Lu, and Xue]{Huang2012}
Huang,~J.,~D.and~Pu; Lu,~Z.; Xue,~Q. \emph{Surface and Interface Analysis}
  \textbf{2012}, \emph{44}, 837--843\relax
\mciteBstWouldAddEndPuncttrue
\mciteSetBstMidEndSepPunct{\mcitedefaultmidpunct}
{\mcitedefaultendpunct}{\mcitedefaultseppunct}\relax
\EndOfBibitem
\end{mcitethebibliography}
\end{document}